# HOW TO CHARACTERIZE HABITABLE WORLDS AND SIGNS OF LIFE

Lisa Kaltenegger

Department of Astronomy and Carl Sagan Institute, Cornell University, Ithaca, New York 14853; email: lkaltenegger@astro.cornell.edu

**ABSTRACT**

The detection of exoplanets orbiting other stars has revolutionized our view of the cosmos. First results suggest that it is teeming with a fascinating diversity of rocky planets, including those in the habitable zone. Even our closest star, Proxima Centauri, harbors a small planet in its habitable zone, Proxima b. With the next generation of telescopes, we will be able to peer into the atmospheres of rocky planets and get a glimpse into other worlds. Using our own planet and its wide range of biota as a Rosetta stone, we explore how we could detect habitability and signs of life on exoplanets over interstellar distances. Current telescopes are not yet powerful enough to characterize habitable exoplanets, but the next generation of telescopes that is already being built will have the capabilities to characterize close-by habitable worlds. The discussion on what makes a planet a habitat and how to detect signs of life is lively. This review will show the latest results, the challenges of how to identify and characterize such habitable worlds, and how near-future telescopes will revolutionize the field. For the first time in human history, we have developed the technology to detect potential habitable worlds. Finding thousands of exoplanets has taken the field of comparative planetology beyond the Solar System.

**CONTENTS**

# 1. COMPARATIVE PLANETOLOGY: LATEST RESULTS

A little more than 20 years since the first extrasolar planets orbiting a Sun-like star were detected (e.g., Latham et al. 1989, Mayor & Queloz 1995, Charbonneau et al. 2000), several thousand exoplanets now provide a first glimpse of the diversity of other worlds (e.g., reviewed in Udry & Santos 2007, Winn & Fabrycky 2015). In the last decade, the field of exoplanet research has transitioned from the detection of exoplanets to characterization of extrasolar giant planets (EGPs; see, e.g., Seager & Deming 2010, Burrows 2014, Crossfield 2015) as well as the discovery of dozens of small exoplanets, which could potentially be habitable (see, e.g., Batalha 2014, Kane et al. 2016). Our closest star, Proxima Centauri, a cool M5V dwarf only 1.3 pc from the Sun, harbors a planet in its habitable zone with a minimum mass of 1.3 Earth masses ($M_\oplus$) that is receiving about 65% of Earth's solar flux (Anglada-Escude et al. 2016). The close-by TRAPPIST-1 planetary system of seven transiting Earth-sized planets around a coolM9Vdwarf star has several (three to four) Earth-size planets in its habitable zone only about 12 pc from the Sun (Gillon et al. 2017). These two planetary systems already show several interesting close-by targets for potentially habitable worlds.

However, a planet is a very faint, small object close to a very bright and large object, its parent star. A comprehensive suite of tools will be needed to characterize habitable planets and moons as the mere detection of a rocky body in the habitable zone (HZ) does not guarantee that the planet is habitable. It is relatively straightforward to remotely ascertain that Earth is a habitable planet, replete with oceans, a greenhouse atmosphere, global geochemical cycles, and life—if one has data with arbitrarily high signal-to-noise ratio (SNR) and spatial and spectral resolution. The interpretation of observations of exoplanets with limited SNR and spectral resolution as well as no spatial resolution, as envisioned for the first-generation instruments, will be far more challenging and implies that we will need to gather information on a planet's environment to understand what we will see at different wavelengths. Encoded in the planet's emergent and transmission spectra is information on the chemical makeup of a planet's atmosphere, and if the atmosphere is transparent, the emergent spectrum also carries some information about surface properties. That makes light a crucial tool to characterize the planet. The presence or absence of spectral features will indicate similarities or differences between the atmospheres of terrestrial exoplanets and that of Earth, and their astrobiological potential.

This review discusses how to identify a potentially habitable rocky planet, how to model and read a planet's spectrum to assess its habitability, and how to search for the signatures of a biosphere. It explores what the best targets for the search for life are, based on our current knowledge and what the next exciting steps in this search are expected to be in the near future. This new field of exoplanet characterization and comparative planetology has shown an extraordinary capacity to combine research in astrophysics, chemistry, biology, and Earth and planetary science and geophysics in a new and exciting interdisciplinary approach to understanding our place in the Universe and to setting planet formation, evolution, and our planet into an overall context.

## 1.1. How to Identify Rocky Planets: Is There a Mass and Radius that Divides the Population of Rocky Planets from That of Gas Planets?

The two most successful detection methods to date are the transit method and radial velocity searches. In the transit method, the planet blocks part of the starlight from our view, thus measuring the planet's area. The radial velocity method measures the minimum mass of a planet (see, e.g., online databases like *https://www.exoplanets.eu*). A transiting exoplanet can be detected by looking for the periodic dimming of the host star's light. The dimming of the star corresponds to the stellar area the planet obscures from our view, thus letting observers derive the radius of the planet as a fraction of the stellar radius.

Radial velocity searches use the radial velocity shifts that a planet induces in the light of its host star, due to the star's movement around the common center of mass of the system, to derive a planet's mass. Without knowing the inclination angle of the planet's orbit, the derived mass from the movement of its host star represents only the minimum mass of a detected planet. The derived mass through radial velocity of a transiting planet corresponds to its real mass because the inclination of its orbit is then known. Even if radial velocity measurements cannot be made, the mass of exoplanets in planetary systems can also be estimated through variations in transit timing, when the gravitational pull of one planet influences the transit period of another planet or through orbital stability constraints for the system (e.g., reviewed in Winn & Fabrycky 2015).

If both mass and radius are known (**Figure 1**), the mean density of the planet can be used to derive its composition and compare it to planets in our own Solar System. However, for most of the thousands of detected exoplanets, we only know the mass or the radius, depending on the detection method. For planets detected with radial velocity, when only the minimum planetary mass is known, planets with minimum masses below 10 $M_\oplus$ are commonly considered rocky; planets with minimum masses above 10 $M_\oplus$ are considered gas planets. If only the radius is known, transiting planets with radii below 2 Earth radii ($R_\oplus$) are commonly considered rocky. For the subsample of 105 small, low-mass exoplanets, both radius and mass are known, which allows us to explore their composition and test these assumptions (**Figure 1**).

**Figure 1** shows the diversity of known exoplanets for planets below 4 $R_\oplus$ and 30 $M_\oplus$. This selection encompasses all exoplanet data for planets with radii below $2R_\oplus$ and $10M_\oplus$, including the error bars on the measurements. Planetary properties are derived in terms of the stellar properties (see, e.g., Kaltenegger & Sasselov 2011, Gaidos 2013, Dressing & Charbonneau 2015, Kane et al. 2016); therefore, the large error bars on the available data sets also show how important the characterization of the host star is for the characterization of planets. Colored lines in **Figure 1** show exoplanetary density models for different compositions from iron (100% Fe) to Earth-like [$MgSiO_3$ (rock)] to a pure (100%) $H_2O$ composition, encompassing the densest to lightest rocky composition for an exoplanet (following Zeng et al. 2016).

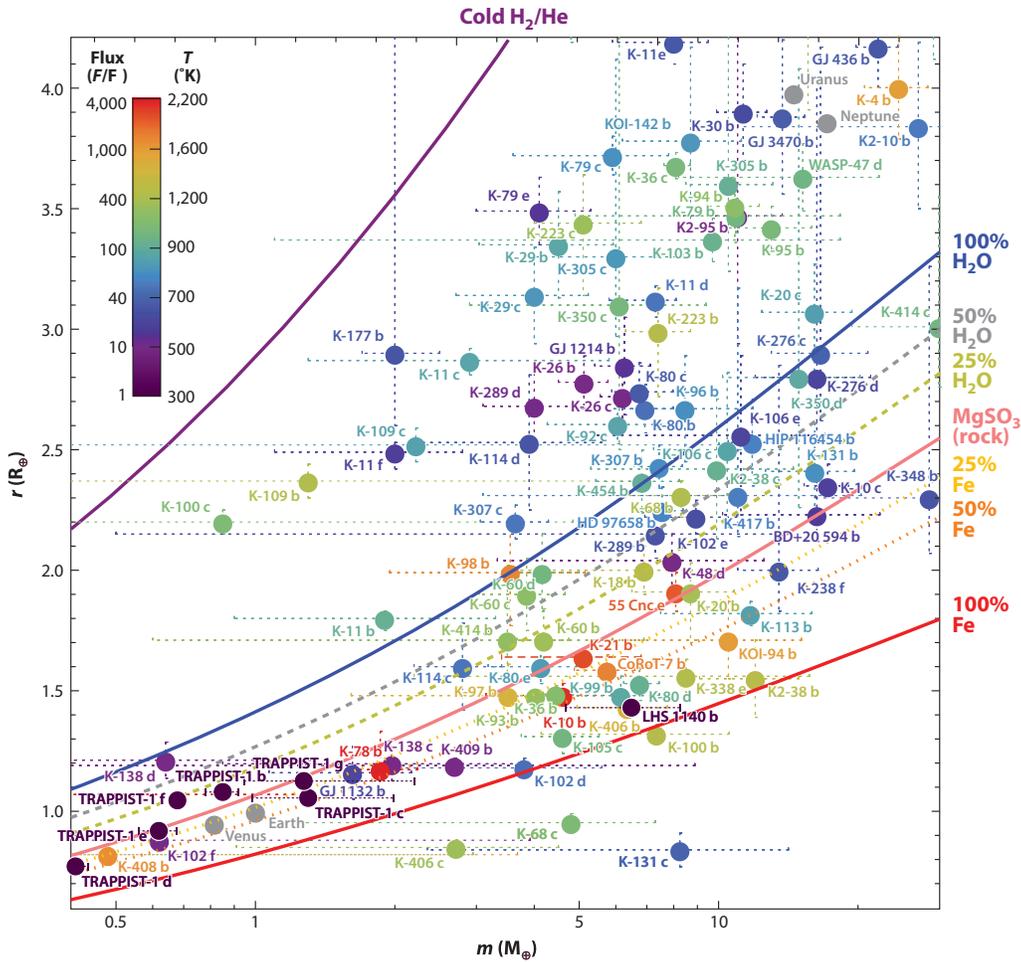

**Figure 1** Mass-radius curves of planets with radii below 4 Earth radii and masses below 30 Earth masses. Planets are color-coded by the stellar flux they receive (compared with Earth). Hypothetical temperatures for the planets are included to add a common physical entity to the diagram and are calculated from the stellar flux received by the planets, assuming a bond albedo of 0, perfect heat redistribution, and no greenhouse effect (e.g., this is a fair estimate for Earth's average surface temperature but not for Venus). Data are from http://www.exoplanet.eu (accessed February 2017) and models following Zeng et al. (2016). Figure courtesy of L. Zeng.

Several known exoplanets with masses below $10 M_\oplus$ have radii corresponding to gas planets or so-called mini-Neptunes; for example, Kepler-11f has a mass between 1.1 and $5 M_\oplus$ but a radius of 2.6 $R_\oplus$ (Lissauer et al. 2011). Another group of exoplanets shown in **Figure 1** consists of so-called super-Earths like Kepler-10c, a planet with approximately 18 $M_\oplus$ and 2.3 $R_\oplus$ (Dumusque et al. 2012), consistent with a rocky composition (see Zeng & Sasselov 2013). Gas planets can have masses down to $2 M_\oplus$, and planets with masses above $10 M_\oplus$ can also be rocky, making the mass of a planet a very weak constraint on its composition. However, so far no planets have been detected that are smaller than approximately 2 $R_\oplus$ and do not fit a rocky composition—that is, that fall in the subset of known density planets shown in **Figure 1**, assuming the nominal values. Therefore, if only one of the two parameters, radius or mass, is available, radius currently sets a stronger constraint on whether the planet is a rocky world or not.

Deriving the limit-dividing rocky and gaseous planets from a subset of the currently available data has been attempted (e.g., Weiss & Marcy 2014). The classic technique of fitting a power law to a noisy data set to derive such a limit has severe limitations in this case because it does not properly account for parameters like measurement uncertainty in planetary radii and mass, non-detections, and upper limits (see Wolfgang et al. 2016). Using a subsample of 27 detected exoplanets and a range of interior planet models, Rogers (2015) derived a high probability of planets being rocky with radii below 1.53 to 1.92 $R_\oplus$, consistent with the data shown in **Figure 1**.

There is no consistent limit in the literature for the mass or radius divide between the terms mini-Neptune and super-Earth, which often leads to confusion, especially for planets that need a substantial gaseous envelope to fit their radius but have masses below 10 $M_\oplus$: These planets are often called super-Earths based only on the 10 $M_\oplus$ limit, for example, GJ1214 (e.g., Zeng & Sasselov 2013) (**Figure 1**).

For the purpose of this review, super-Earths are rocky planets with outgassed atmospheres and a mean density higher than that of water, and mini-Neptunes are planets that have a substantial primordial gaseous envelope and a mean density lower than that of water (a division that is indicated by the 100% $H_2O$ line in **Figure 1**). Observationally, their mass and radius distinguish such planets. The observable spectra of mini-Neptunes with hydrogen atmospheres should also differ strongly from the spectra of rocky planets with outgassed atmospheres like Earth due to the difference of the atmospheric makeup.

## 1.2. How to Use Stellar Incident Flux to Identify Interesting Exoplanets

The color-coding in **Figure 1** shows the stellar flux the detected exoplanets receive, allowing us to compare the stellar irradiance and to trace any high irradiance and thus high temperature effect on the planetary population as well as identify interesting exoplanets for follow-up observations. **Figure 1** shows extremely highly irradiated rocky planets, like Kepler-10b, that are very interesting planets for future observations. They are, of course, not considered as habitable worlds, but assuming that their surface is molten at these extremely high stellar irradiances, their atmospheric makeup should showcase the material composition of the planet's crust (see, e.g., Schaefer & Fegley 2009, Miguel et al. 2011). Observing such highly irradiated worlds could give us clues to the composition and formation of rocky exoplanets (see, e.g., Williams & Cieza 2011).

This stellar incident flux is also translated in **Figure 1** into an approximate equilibrium or equivalent blackbody temperature, which expresses a theoretical temperature that a planet would be at if it were a blackbody being heated only by its parent star, assuming a planetary albedo of zero and no greenhouse effect. However, the equilibrium temperature of a planet gives almost no information about the surface temperature of a planet with a substantial atmosphere. Deriving surface temperatures for planets with atmospheres requires detailed measurements of their atmosphere's chemical makeup, which is not yet available for rocky exoplanets, except for single cases of highly irradiated worlds like 55 Cancri e (Demory et al. 2016). Therefore, temperature values for potentially habitable rocky planets in the literature need to be evaluated critically.

However, one can use the incident stellar flux planets receive to compare planetary environments: Present-day Venus, for example, receives 1.9 times the solar flux at Earth's orbit, $S_0$, and present-day Mars receives 0.4 $S_0$. Any rocky planet that receives more flux than present-day Venus is empirically too hot to be habitable. Note that the different assumptions on how high and how low the stellar irradiance is that still allows for habitability have led to a wide range of different values for the abundance of habitable planets in the literature, as shown in **Table 1**.

## 2. WHERE TO LOOK FOR HABITABLE WORLDS

In situ sampling for exoplanets is not possible; therefore, signatures of life need to modify the atmosphere or surface of a planet to be remotely observable. This limits remote detectability of life to planets and moons that can sustain liquid water on their surface, because it remains to be demonstrated whether subsurface biospheres can modify a planet's surface or atmosphere in detectable ways.

**Table 1** $\eta_\oplus$ values: Frequency of rocky planets within the Habitable Zone

| Value | Radius ($R_\oplus$) or mass ($M_\oplus$) | Reference | Notes |
|---|---|---|---|
| 0.01–0.03 | $0.8 \leq r \leq 2.0$ | Catanzarite & Shao 2011 | Host stars: F, G, K; HZ: 0.75–1.8 AU for a solar twin |
| $0.34 \pm 0.14$ | $0.5 \leq r \leq 2.0$ | Traub 2011 | Host stars: F, G, K; HZ: empirical, nominal, narrow (Kopparapu et al. 2014) |
| $0.41^{+0.54}_{-0.13}$ | Mass range: $1 \leq m^*\sin(i) \leq 10\ M_\oplus$ | Bonfils et al. 2013 | Host stars: RV M dwarfs; HZ: Selsis et al. 2007 |
| 0.46 (95% confidence interval: 0.31–0.64) | $0.8 \leq r \leq 2.0$ | Gaidos 2013 | Host stars: Kepler dwarfs; HZ: Selsis et al. 2007 |
| $0.15^{+0.13}_{-0.06}$ | $0.5 \leq r \leq 1.4$ | Dressing & Charbonneau 2013 | Host stars: cool stars ($T < 4{,}000$ K); Narrow HZ (Kasting et al. 1993) |
| $0.48^{+0.12}_{-0.24}$ (conservative HZ); $0.53^{+0.08}_{-0.17}$ (optimistic HZ) | $0.5 \leq r \leq 1.4$ | Kopparapu 2013 | Host stars: M dwarfs; HZ limits (Kopparapu et al. 2013) |
| $0.22 \pm 0.08$ | $1.0 \leq r \leq 2.0$ | Petigura et al. 2013 | Host stars: solar type, HZ: 0.25–4.0 times flux on Earth ($F_\oplus$), twice as much flux as on Venus |
| 0.25 | $0 < r < 1.4$ | Morton & Swift 2014 | Host stars: cool dwarfs ($T < 4{,}000$ K); Construct planet radius distribution to small radii, using M-star planets from the Kepler data available ca. 2014 |
| $0.06^{+0.03}_{-0.01}$ | $1.0 \leq r \leq 2.0$ | Silburt et al. 2015 | Host stars: solar type; HZ: 0.99–1.7 AU |
| $0.16^{+0.17}_{-0.07}$ (Earth-sized); $0.12^{+0.10}_{-0.05}$ (super-Earth-sized) | $1.0 < r < 1.5$; $1.5 < r < 2.0$ | Dressing & Charbonneau 2015 | Host stars: M dwarfs (update on their 2013); Conservative HZ (moist greenhouse/maximum greenhouse) |
| $0.24^{+0.18}_{-0.08}$ (Earth-sized); $0.21^{+0.11}_{-0.06}$ (super-Earth-sized) | $1.0 < r < 1.5$; $1.5 < r < 2.0$ | Dressing & Charbonneau 2015 | Host stars: M dwarfs (update on their 2013 paper); Optimistic HZ (RV/early Mars) |
| 0–0.3 | $0.5 \leq r \leq 2.0$ | Zsom 2015 | HZ: range of fluxes $\sim$2–3 $F_\oplus$ for inner edge, $\sim 10^{-2}$ $F_\oplus$ for outer edge |
| F: (a) $0.59 \pm 0.12$, (b) $0.66 \pm 0.14$[a]<br>G: (a) $0.97 \pm 0.02$, (b) $1.03 \pm 0.10$[a]<br>K: (a) $0.72 \pm 0.02$, (b) $0.75 \pm 0.11$[a]<br>M: (a) $0.75 \pm 0.33$, (b) $1.23 \pm 0.18$[a] | $0.5 \leq r \leq 1.25$ | Traub 2016 | Using a new method for inferring the exoplanet population from Kepler data, based on assuming the planet frequency can be represented by a smooth function of planet radius and period; HZ: 0.8–1.8 AU for a solar twin |

[a] The *a* and *b* values were derived from two different *Kepler* data releases: (*a*) q1-16 and (*b*) q1-17.
Abbreviations: HZ, habitable zone; RV, recent Venus.

Our search for signs of life is also based on the assumption that extraterrestrial life shares fundamental characteristics with life on Earth in that it requires liquid water as a solvent and has a carbon-based chemistry (for a wider range of possibilities, see, e.g., Bains 2004, Chyba & Hand 2005, Baross et al. 2007, Brack et al. 2010). Life based on a different chemistry is not considered, because how such life could function as well as what signatures it could produce in the atmosphere or the surface of a planet is so far unknown. The idea of habitability and thoughts on what limits and facilitates it on planets and moons has been discussed in detail in several reviews (see, e.g., Baross et al. 2007, Southam & Westall 2007, Lineweaver & Chopra 2012, Cockell 2016).

Earth serves as a Rosetta stone in the search for habitats. Earth's atmosphere has changed significantly through its evolution (e.g., review in Kasting & Catling 2003). Even though we cannot observe Earth at different geological ages, we can use rock records to constrain the atmospheric makeup of our planet through time (see detailed discussion below). Due to the known evolution of the Sun, we can also constrain the stellar irradiance on Earth as well as other planets in the Solar System through geological time. For Earth, the combination allows us to model remotely detectable spectral surface signatures, and signatures of life through geological time (see, e.g., Kaltenegger et al. 2007, Grenfell et al. 2010). The surface ultraviolet (UV) environments on our planet through geological time can also be constrained through modeling the UV flux that would reach the surface at different geological epochs (Segura et al. 2005, Rugheimer et al. 2015b, Arney et al. 2016).

The key to identifying which exoplanets could be potential habitats and prioritizing time intensive observations consists of identifying important parameters that allow for life on Earth and using models to export those parameters to other potential habitats. Signatures of life on Earth are also our first insight into which signs of biospheres could be remotely detectable on other planets in different viewing geometries (see, e.g., Lovelock 1965, Sagan et al. 1993, Des Marais et al. 2002, Kaltenegger et al. 2009, Seager et al. 2016).

## 2.1. How to Model a Habitable Planet

A planet's climate is mainly influenced by two factors: the stellar irradiance and the atmospheric composition. Here, we assume that ocean and atmospheric dynamics allow for effective heat distribution on the planet. If that is not the case, the dynamics of a planet's atmosphere are also a crucial component of its climate and need to be addressed in three-dimensional (3D) general circulation models (GCMs), as discussed below. The atmospheric makeup of a planet depends on the sources and sinks of the gases that make up its atmosphere as well as the photochemical reactions in the atmosphere. The sources and sinks of different chemicals depend on the composition of the planet's crust, whether it has an ocean or not, geochemical cycles, geological activity, and if biota exists, its biological cycles. Which geochemical cycles exist and which one dominates on a planet depends on the planetary environment: On Earth, the carbonate-silicate cycle dominates the long-term climatic stability of the planet, whereas on dry planets, for example, the $SO_2$ cycle could dominate the climate, leading to a very different atmospheric composition for similar outgassing rates, crust material, and stellar irradiance (see, e.g., Kaltenegger & Sasselov 2010 for details).

A planet is warmed by absorption of stellar radiation of visible and near-infrared (NIR) radiation from its host star (and potentially internal heating) and cooled by emission of thermal IR radiation. If the planet were a blackbody, we could easily derive its equilibrium temperature $T_{equ}$ using the energy balance (Equation 1)

$$\delta\, T_{equ}^4 = S/4\, (1-A) \qquad (1)$$

where $\delta$ is the Stefan-Boltzmann constant ($5.67 \times 10^{-8}$ W/m$^2$/K$^4$), S the stellar irradiance at the planet's orbit and A the reflectivity or albedo of the planet. If we apply this equation to Earth (using A=0.3, $S_0$=1370W/m$^2$) we derive $T_{equ\_Earth}$ = 255K. The 33-K difference compared to Earth's average surface temperature ($T_{surf}$) of 288 K is due to Earth's atmosphere. Earth's IR emission is reabsorbed and re-emitted by IR-active gases in the atmosphere, leading to a greenhouse effect of planets with IR-active gases in their atmosphere. For present-day Earth, the two most important greenhouse gases are $H_2O$ (approximately two-thirds of the greenhouse effect) and $CO_2$ (approximately one-third of the greenhouse effect), and lesser contributions are from methane, nitrous oxide, and ozone ($CH_4$, $N_2O$, $O_3$) and various chlorofluorocarbons (CFCs) (reviewed in Kasting & Catling 2003).

The three rocky planets in our Solar System with substantial atmospheres, Venus, Earth, and Mars, have very different atmospheric compositions: Their surface pressures are 92 bar, 1 bar, and 6.36 mbar, respectively. Their Bond albedos, that is, the reflectivity of a planet integrated over the entire wavelength range, are 0.77, 0.3, and 0.25, respectively. Their mean surface temperatures are approximately 460°C, 15°C, and −55°C, respectively[1]. Taking their respective albedo values and comparing them to the planets' $T_{surf}$ shows that the greenhouse effect accounts for 33 K of warming for Earth, 523 K for Venus, and 10 K for Mars. The denser the atmosphere, the worse $T_{equ}$ characterizes the planet. Although $T_{equ}$ is a useful quantity for blackbodies, this comparison of the Solar System rocky planets shows that it gives almost no information about any surface temperature of a planet with a substantial atmosphere. That shows that remote observations of the chemical makeup of an atmosphere are critical to derive surface conditions.

Venus, Earth, and Mars show the wide range of albedos of the rocky planets in our own Solar System. On Venus and Earth, most of the albedo is due to highly reflective clouds. Clouds are observed in Earth's atmosphere and parameterized. For planets with conditions similar to those in Earth's atmosphere, water clouds should show similar characteristics. However, for very different conditions, unknown cloud properties and cloud feedback responses lead to large uncertainty in the models. Exploring the effect of cloud parameters like particle size distribution and density on a planet's climate has given us the first insights on the influence of clouds on a planet's climate and albedo (see, e.g., Kitzmann 2017, Mischna et al. 2000, Zsom et al. 2012) and remotely detectable spectral features (see, e.g., Kaltenegger et al. 2007, Rauer et al. 2011).

This simple comparison shows the diversity of rocky planets in our own Solar System, which gives us insights into their evolution. That evolution can be explained through their distance to the Sun and subsequent evolution (e.g., reviewed in Kasting & Catling 2003). The stellar irradiance is a useful characteristic of a planet's environment because a planet's climate is shaped by stellar irradiance and climate feedbacks.

**2.1.1. Climate feedback on Earth.** First, we will look at climate feedbacks: Positive feedbacks further increase warming (or cooling) due to temperature increase (or decrease), whereas negative feedbacks decrease such warming (or cooling), stabilizing the climate system (see Pierrehumbert 2010).

Water is near its condensation temperature on Earth and the water vapor feedback is nearly instantaneous. When the climate cools, the atmospheric water vapor decreases proportionally, leading to a smaller greenhouse effect due to $H_2O$, cooling the planet further. When the climate warms, atmospheric water vapor increases and warms the planet further due to the greenhouse effect. An increase in surface temperature decreases the snow and ice cover, reducing Earth's albedo, further warming the planet, whereas a decrease in surface temperature increases snow and ice cover, increasing the albedo and cooling the planet further in what is called the snow/ice albedo feedback. If $T_{surf}$ increases, the outgoing IR emission of a planet also increases, cooling the planet and reducing $T_{surf}$, thereby stabilizing Earth's climate on short timescales.

On long timescales, the negative feedback loop that stabilizes climate is the $CO_2$ climate feedback, which also allows for habitability on rocky planets over geological time. The inorganic part of the carbon cycle, the carbonate-silicate cycle (**Figure 2**), stabilizes Earth's long-term climate (Walker et al. 1981). $CO_2$ dissolves in rain water to carbonic acid that over long timescales dissolves silicate rocks. The products of silicate weathering are transferred to the oceans and used by organisms to make shells of calcium carbonate or silica. When the organisms die and fall to the seafloor, most of the shells dissolve, but a fraction is buried in seafloor sediments, reducing atmospheric $CO_2$. Because the seafloor is created at mid-ocean ridges and subducted at certain plate boundaries, where the denser ocean plates subduct under the lighter continental plate, the sediments are transported to higher temperatures and pressures, reforming and in the process releasing $CO_2$ that enters the atmosphere again through volcanism.

---

[1] http://nssdc.gsfc.nasa.gov/planetary/planetfact.html

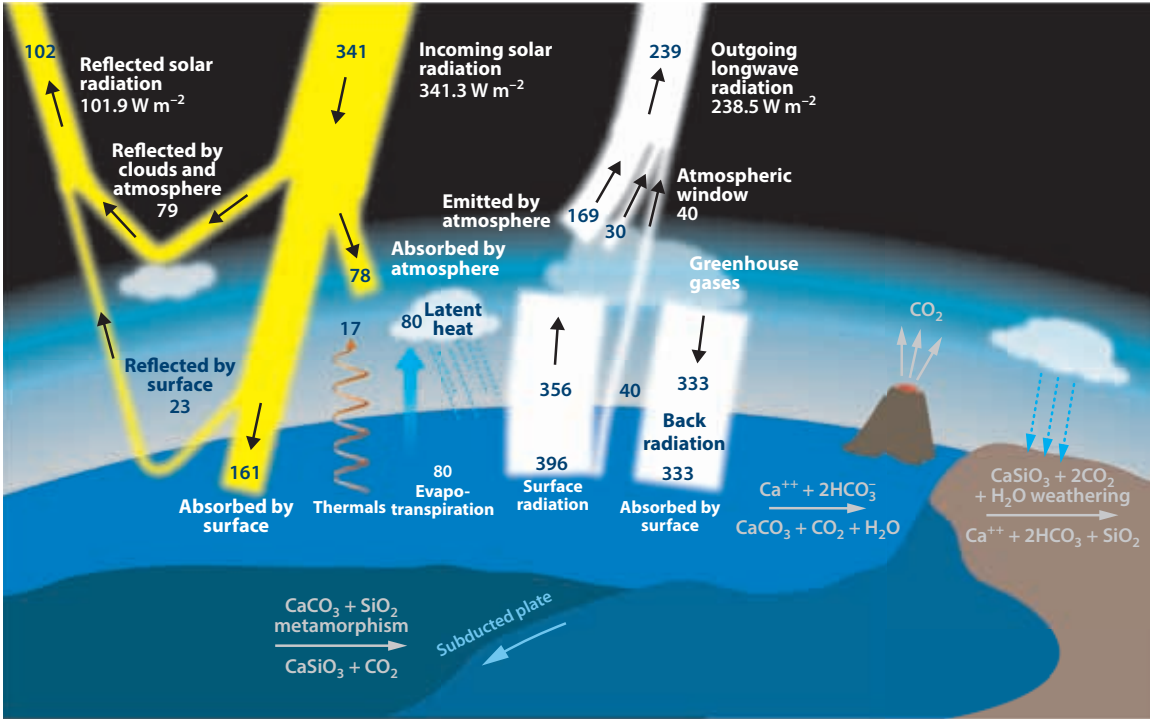

**Figure 2** Estimate of Earth's annual and global mean energy balance, and sketch of the modern carbonate-silicate cycle. Data from Trenbeth et al. (2009) and Kasting & Catling (2003) respectively. Note that 342 W/m$^2$ represents the Solar Constant 1368 W/m$^2$ divided by 4, accounting for day and night as well as average incident angle of solar irradiation (Trenbeth et al 2009).

The timescale of this cycle is hundreds of thousands to a million years and is sensitive to $T_{surf}$ (see, e.g., Walker et al. 1981, Berner et al. 1983). With increasing $T_{surf}$, both the chemical reaction rate and the evaporation and precipitation increase, in turn reducing atmospheric $CO_2$ and cooling $T_{surf}$. The carbon cycle on Earth also has an organic component that acts over shorter timescales during which plants and microbes convert $CO_2$ and $H_2O$ into organic matter and release $O_2$ by photosynthesis. Respiration and decay balance this process by generating $CO_2$ and $H_2O$..

Because of this balance, the net release of $O_2$ in the atmosphere is from the burial of organics in sediments (**Figure 3**).

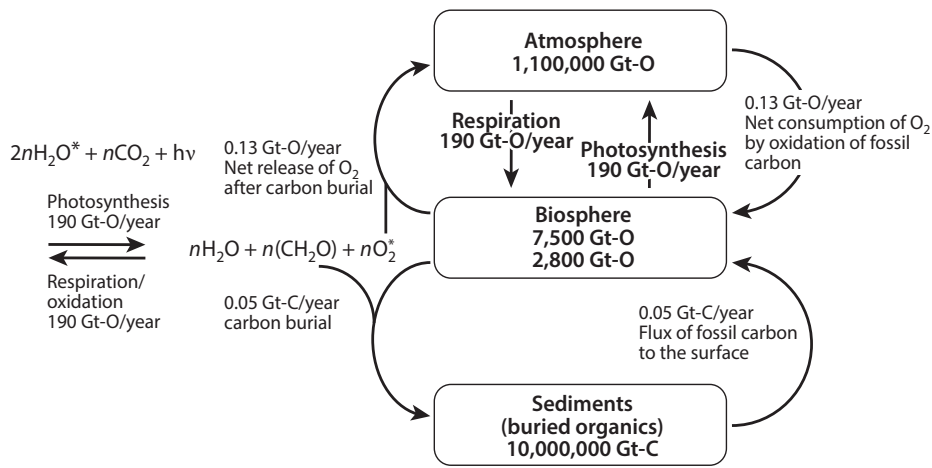

**Figure 3** The oxygen cycle on Earth (Kaltenegger et al. 2009). Figure courtesy of F. Selsis.

Each reduced carbon buried results in a free $O_2$ molecule in the atmosphere. This net release rate is also balanced by weathering of fossilized carbon when exposed to the surface. The oxidation of reduced volcanic gases such as $H_2$ and $H_2S$ also accounts for a significant fraction of the oxygen losses. The atmospheric oxygen is recycled through respiration and photosynthesis in less than 10,000 years. In the case of a total extinction of Earth's biosphere, the atmospheric $O_2$ would disappear in a few million years (see, e.g., Kaltenegger et al. 2009).

**2.1.2. Earth's global mean energy balance.** A detailed view of Earth's estimated annual and global mean energy balance (in W m−2) is shown in **Figure 2**. The incoming solar radiation at the top of the atmosphere is approximately 342 W m−2, with roughly 30% reflected to space. Seventy percent of the reflection is due mainly to clouds and aerosols, with Earth's surface responsible for the remainder. The surface absorbs most of the remaining solar radiation; thus, atmospheric warming occurs from below, establishing the vertical temperature profile and the large-scale circulation of the atmosphere. The outgoing longwave radiation that balances Earth's energy budget is mainly emitted by the atmosphere and clouds. The large majority of surface longwave radiation is absorbed by IR-active atmospheric greenhouse gases and re-emitted toward the surface. The net mass and heat transfer from the surface to the atmosphere, one of the main drivers of the general atmospheric circulation, is due to evaporation or sublimation at the surface, cooling the surface, and the formation of clouds, where latent heat is released in the atmosphere.

The balance of in- and outgoing fluxes guides any atmospheric model that characterizes Earthlike planets. For extrasolar planets or moons, internal heating could also add to the warming of a planet's surface due to radioactive decay or tidal heating, depending on the orbital configuration of the system (see, e.g., Barnes et al. 2009, Henning et al. 2009, Henning & Hurford 2014, Dobos & Turner 2015). How much warming one would expect for different cases depends on the complexity of interior models used and is still debated.

To model a wide range of rocky planets and habitable worlds, single-column models use a plane-parallel approach that divides the atmosphere in layers (see, e.g., Kasting et al. 1993, Segura et al. 2005, Tinetti et al. 2006, Selsis et al. 2007, Kaltenegger et al. 2009, Rauer et al. 2011, Kopparapu et al. 2013, Rugheimer et al. 2013, Domagal-Goldman et al. 2014, Ramirez et al. 2014a,b). Potentially habitable exoplanets will be unresolved and therefore present a disk-integrated spectrum. Models of disk-integrated spectra match observations obtained when using the Earth's average atmosphere profile (see, e.g., Woolf et al. 2002, Montanes-Rodr´ıguez et al. 2006, Turnbull et al. 2006, Kaltenegger et al. 2007, Arnold 2008, Robinson et al. 2011, Palle et al. 2016).

**2.2. WHERE TO FIND A HABITABLE PLANET: THE HABITABLE ZONE**
The HZ is used to identify potentially habitable planets and guide remote detection of life. It is a tool that prioritizes rocky planets for follow-up observations. The HZ is the circular region around one or multiple stars in which liquid water could be stable on a rocky planet's surface (e.g., Kasting et al. 1993, Haghighipour & Kaltenegger 2013, Kaltenegger & Haghighipour 2013, Kane & Hinkel 2013), facilitating the detection of possible atmospheric biosignatures. The HZ concept was proposed for the first time by Huang (1959) and has since been calculated by several authors (e.g., Hart 1978, Kasting et al. 1993, Selsis et al. 2007, Abe et al. 2011, Pierrehumbert & Gaidos 2011, Kopparapu et al. 2013, Cullum et al. 2014, 2016, Read 2014, Ramirez & Kaltenegger 2014, 2017).

It is important to note that the HZ is not the region around a star where life is possible, or exists, but the region around a star where liquid water is possible on the surface of a geologically active rocky planet. Liquid surface water is used because it remains to be demonstrated whether subsurface biospheres, for example, under an ice layer on a frozen planet, can modify a planet's atmosphere in ways that can be detected remotely. The liquid water HZ for Earth-like planets (see, e.g., Kasting & Catling 2003) would be a more accurate nomenclature, but most people shorten it to habitable zone. The shorter version can indeed be misleading without knowing the context and purpose of the HZ.

The width and distance of a given HZ depends to a first approximation on two main parameters: incident stellar flux and planetary atmospheric composition. The incident stellar flux depends on stellar luminosity, stellar spectral energy distribution, eccentricity of the system, and the planet's orbital distance. The warming due to atmospheric composition depends on the planet's atmospheric makeup, energy distribution, and resulting albedo and greenhouse warming. In the literature, very different values of stellar irradiance are used as boundaries for the HZ (**Table 1**). We discuss the different limits and their constraints below.

**2.2.1. Deriving limits for the Habitable Zone.** The empirical HZ is based on observations in our own Solar System (see Kasting et al. 1993). The inner edge of this empirical HZ, the so-called recent Venus (RV) limit, is based on the observation that Venus may have had liquid water on its surface until approximately 1 Ga, consistent with atmospheric D/H ratio measurements suggesting a high initial water endowment (e.g., Donahue et al. 1982). At that time the Sun was approximately 8% less bright than today, yielding a solar flux equivalent equal to 1.76 that of present-day solar irradiance at Earth's orbit (S0). The empirical outer edge for the HZ, the so-called early Mars limit, is based on observations suggesting that Mars did not have liquid water on its surface at or after 3.8 Ga. At that time the solar flux was approximately 25% lower or equal to approximately 0.32 S0. The corresponding orbital distances in our Solar System are 0.75 AU (RV limit) and 1.77 AU (early Mars limit), respectively, for present solar luminosity, excluding present-day Venus and including present-day Mars.

The stellar energy distribution (SED) changes for stars of different spectral types and ages. A star's radiation shifts to longer wavelengths with cooler surface temperatures, which makes the light of a cooler star more efficient in heating an Earth-like planet with a mostly $N_2$- $H_2O$ -$CO_2$ atmosphere. This is partly due to the effectiveness of Rayleigh scattering, which decreases at longer wavelengths. A second effect is the increase in NIR absorption by $H_2O$ and $CO_2$ as the star's spectral peak shifts to these wavelengths, meaning that the same integrated stellar flux that hits the top of a planet's atmosphere from a cool red star warms a planet more efficiently than the same integrated flux from a hot blue star. Stellar luminosity as well as SED changes with spectral type and stellar age, which influences the orbital distance at which a planet can maintain Earth-like climate conditions or, more generally, liquid water on its surface.

**2.2.2. The Classical Habitable Zone.** In contrast to the observation-based empirical HZ, a narrower classical HZ can be derived using 1D atmospheric models (Kasting et al. 1993) for main sequence (MS) stars with effective temperatures ($T_{eff}$) between 2,600 K and 7,200 K (Kopparapu et al. 2013, 2014, Ramirez et al. 2014a,b) and up to 10,000 K (Ramirez & Kaltenegger 2016) (**Table 2**).

**Table 2 Constants to compute the empirical main sequence (MS) and post-MS Habitable Zone boundaries using Equation 1[a]**

| Constant | Recent Venus limit: inner edge | Three-dimensional model limit: inner edge | Early Mars limit: outer edge |
|---|---|---|---|
| $S_\odot$ | 1.7665 | 1.1066 | 0.324 |
| A | $1.3351 \times 10^{-4}$ | $1.2181 \times 10^{-4}$ | $5.3221 \times 10^{-5}$ |
| B | $3.1515 \times 10^{-9}$ | $1.534 \times 10^{-8}$ | $1.4288 \times 10^{-9}$ |
| C | $-3.3488 \times 10^{-12}$ | $-1.5018 \times 10^{-12}$ | $-1.1049 \times 10^{-12}$ |

[a]Data taken from Kopparapu et al. (2014) and Ramirez & Kaltenegger (2014).

The classical HZ is based on specific assumptions: We assume other rocky planets are geologically active and regulate the $CO_2$ in their $N_2$- $CO_2$- $H_2O$ atmospheres via geochemical cycles. The classical HZ is defined by the greenhouse effect of two gases: $CO_2$ and $H_2O$ vapor. The outer edge of the HZ is defined as the distance beyond which condensation and scattering by $CO_2$ outstrip its greenhouse capacity, the so-called maximum greenhouse limit of $CO_2$. The inner edge occurs when mean surface temperatures exceed the critical point of water, triggering a runaway greenhouse state that leads to rapid water loss to space on very short timescales. The entire surface water reservoir can be vaporized by runaway greenhouse conditions, followed by the photo-dissociation of water vapor and subsequent escape of free hydrogen into space. For Earth-like planets, the runaway greenhouse state (i.e., complete ocean evaporation) would be triggered either when the planet's surface temperature reached the critical temperature for water (647 K) (see Kasting 1988 for details) or in an atmosphere in which the absorbed solar flux exceeded the outgoing thermal-IR flux (Goldblatt &Watson 2012).

As discussed above, on a geologically active planet like Earth, the geochemical carbonate silicate cycle stabilizes the long-term climate and atmospheric $CO_2$ content, depending on the surface temperature: $CO_2$ is continuously outgassed and in the presence of surface water forms carbonates, which then get subducted and $CO_2$ is again outgassed. Farther from the star, the lower stellar irradiance would create a cooler surface temperature on a planet, thus linking the orbital distance to atmospheric $CO_2$ concentration levels: $CO_2$ should be a trace gas close to the inner edge of the HZ but a major compound in the outer part of the HZ with several bar of $CO_2$ (e.g., Walker et al. 1981). Because the outer limit of the HZ is based on the assumption that atmospheric $CO_2$ will build up and warm the planet, adequate $CO_2$ outgassing rates are needed. For low $CO_2$ outgassing rates, the climate of a planet could repeatedly cycle between unstable glaciated and deglaciated climatic states at the outer edge of the classical HZ (see, e.g., Kadoya & Tajika 2014, Menou 2015, Haqq-Misra et al. 2016).

Close to the inner edge of the HZ for increasing stellar irradiation, the models show that a very dense, water-rich atmosphere would develop because more and more of Earth's oceans would evaporate, creating a steam atmosphere with several hundred bar of pressure. Such changes in the atmospheric makeup of exoplanets linked to their position in the HZ should be detectable with upcoming telescopes.

The 1D limits derived for the classical HZ are based on atmospheric Earth models that do not take cloud feedback response into account, that is, how clouds vary for different atmospheric conditions. Clouds are responsible for most of the planetary albedo on present-day Earth (see, e.g., Kasting & Catling 2003). They reflect incident solar radiation and also absorb surface IR emission. Therefore, clouds are an important input to exoplanet climate models for various conditions like the limits of the HZ. Cloud characteristics as well as coverage are linked to large-scale atmospheric circulation of the atmosphere, which is influenced by several factors, such as rotation rate and incident stellar flux. Three-dimensional models are needed to assess how high clouds and cloud coverage in general would change for very different planetary environments. However, modeling clouds is very complex and no data exist that can provide comparison data sets for, for example, fast-rotating or slow-rotating, very hot or very cold Earth-like planets.

A lively discussion exists in the literature, based on extrapolation of 3D models to different environments, on how water and $CO_2$ clouds as well as rotation rate would change the climate and the limits of the HZ (see, e.g., Forget & Pierrehumbert 1997, Lorenz et al. 1997, Joshi 2003, Williams & Pollard 2002, Lopez et al. 2005, Selsis et al. 2007, Edson et al. 2011, Zsom et al. 2012, Goldblatt et al. 2013, Leconte et al. 2013a,b, 2015, Vladilo et al. 2013, Wordsworth & Pierrehumbert 2014, Yang et al. 2013, 2014, Fereira et al. 2014, Wolf & Toon 2015, Linsenmeier et al. 2015, Kopparapu et al. 2016, Kitzmann 2017, Ramirez & Kaltenegger 2017). Both 1D and 3D models generally indicate wider boundaries when considering cloud feedback than the classical HZ, but model results differ in their specifics. Therefore, in this review, we show the 1D empirical HZ limits and, for comparison, a 3D model limit (based on Leconte et al. 2013a). The outer HZ limits agree in the 3D and 1D models. Surface pressure and gravity between 0.5 and 5 $M_\oplus$ only change the MS HZ limits a few percent (Kopparapu et al. 2014).

Some models suggest that planets with high obliquities may remain habitable farther from their host stars, because the poles would receive large energy fluxes, suppressing the runaway glaciation at the outer edge of the HZ (e.g., William & Kasting 1997, Spiegel et al. 2009). Slow-rotating planets could maintain habitability for higher incident fluxes than fast-rotating planets due to increased cloud coverage on the dayside of the planet, making synchronously locked planets at the inner edge of the HZ interesting test cases for this hypothesis (see, e.g., Wolf & Toon 2014, Yang et al. 2014, Kopparapu et al. 2016).

**2.2.3. New Habitable Zone concepts.** Adding additional greenhouse gases to a $N_2$- $CO_2$- $H_2O$ atmosphere can extend the HZ outward or move it, depending on whether the gas condenses in the atmosphere. Adding, for example, hydrogen to the atmosphere shifts the surface liquid water zone around a star outward; therefore, the exact limits of the HZ depend on the atmospheric makeup of a planet.

Young planets can accrete many bar of primordial hydrogen, which is a potent greenhouse gas that can extend the HZ temporarily outward by several AU: A 40-bar hydrogen atmosphere would extend the liquid water surface temperatures out to 10 AU in our Solar System (Pierrehumbert & Gaidos 2011). For comparison, the greenhouse effect of $CO_2$ is outstripped by the effects of condensation and scattering at high $CO_2$ partial pressures, which limits the outer edge of the HZ to approximately 1.7 AU in our own Solar System (e.g., Kasting et al. 1993).

The high greenhouse efficiency of $H_2$ arises from collision-induced absorption caused by self-broadening from $H_2$- $H_2$ collisions, allowing $H_2$ to function without condensing out to great distances (Pierrehumbert & Gaidos 2011). For planetary atmospheres that are not hydrogen dominated, foreign-broadening of hydrogen by the background atmosphere excites roto-translational bands within the hydrogen that promote significant absorption in spectral regions where $CO_2$ and $H_2O$ absorb poorly, enhancing warming (Ramirez et al. 2014a,b, Wordsworth et al. 2017). However, hydrogen quickly escapes to space. Without a continuous source, hydrodynamic escape would strip a super-Earth HZ planet with 50 bar of primordial atmospheric $H_2$ in a few million years (Wordsworth 2012).

Smaller amounts of hydrogen on terrestrial planets could be continuously supplied through volcanic outgassing. The input of hydrogen from volcanic sources is balanced by its escape to space. Actual hydrogen escape rates will be planet specific, with characteristics such as increasing stellar irradiance increasing and planetary mass decreasing escape rates. Adding approximately 30% of the dry atmosphere of an Earth-mass planet in $H_2$ to a $N_2$- $CO_2$- $H_2O$ atmosphere could be sustained through volcanism and would extend the HZ outward by approximately 30–35% for A to M host stars (Ramirez & Kaltenegger 2017). This volcanic hydrogen HZ ($N_2$- $CO_2$- $H_2O$ - $H_2$) would extend out to 2.4 AU in our Solar System, whereas the inner edge is only moved by 1–6% due to the large amount of water vapor dominating the climate on the inner edge of the HZ.

Adding hydrogen to a planet's atmosphere also decreases the mean molecular weight of the atmosphere, which increases its atmospheric scale height and thereby promotes detectability of atmospheric features for future telescopes. Whether one could identify and detect biosignatures on a planet with a hydrogen atmosphere is an interesting open question (see, e.g., Hu et al. 2012, Kasting et al. 2014, Seager et al. 2013).

Removing a greenhouse gas from a $N_2$- $CO_2$- $H_2O$ atmosphere will also shift the limits of the HZ. Using a 3D model, Abe et al. (2011) showed that the inner limit of the HZ for dry planets moves inward by approximately 10%, depending on how much water is still in the atmosphere. A 3D model calculates the relative humidity of a planet with liquid water on its surface self consistently. Without calculating the relative humidity self-consistently, 1D models have to set this value, normally derived from observations or 3D models. Using a very low relative humidity in a 1D model, Zsom et al. (2014) found that the inner HZ in our Solar System could move much closer to the Sun, to 0.38 AU. However, Kasting et al. (2014) subsequently argued that setting such low relative humidity levels while still assuming liquid water on the planet's surface violates surface energy balance. This case shows how the combination of 1D models that explore the parameter space and 3D models that validate specific assumptions made for non-present-day Earth-like conditions are an important combination to explore the parameter space of extrasolar planets.

**2.2.4. The Binary Habitable Zone.** Exoplanets have also been found orbiting binaries and higher order systems (e.g., reviewed in Winn & Fabrycky 2015). The binary HZ (O'Malley-James et al. 2012, Haghighipour & Kaltenegger 2013, Kaltenegger & Haghighipour 2013, Kane & Hinkel 2013, Cuntz 2014, Forgan 2014) is based on the radiation the planet receives from both stars. In lieu of modeling the heating due to irradiation from two stars on a binary planet, one can calculate where the binary HZ is by comparing the sum of the weighted flux of both stars to the stellar irradiance limits of the HZ for our Sun. As discussed above, stars with cooler surface temperatures warm a planet's surface more efficiently; thus, one cannot just add up the flux from two different star types and compare them to the HZ flux limits in the Solar System. The easiest way to find the orbital distance of the binary HZ is to first compare the efficiency of heating of each binary star to that of our Sun, given by the stellar flux weighting factor for each star type in Kaltenegger & Haghighipour (2013), sum up the weighted stellar flux, and then compare at what orbital distance in the exoplanetary system our Sun's incident stellar flux limits of the HZ are reached. This approach can be used for planets in multiple star systems as well, not only binaries.

If a planet in a binary system orbits both stars, a so-called P-type binary planet, it will always receive flux from both stars, except if they temporarily eclipse each other. Thus, to first order, the binary HZ corresponds to a specific orbital distance as if the planet were only orbiting a single star. However, if the planet only orbits one of the two stars, a so-called S-type binary planet, the orbital distance of the HZ from the host star changes depending on the varying distance of the second star from the planet. In extreme cases, the stellar irradiation from the second binary star could move a planet temporarily out of the binary HZ, making such planets very interesting cases to study how well a planet's atmosphere can buffer temporary changes in overall stellar irradiation.

**2.3. THE HABITABLE ZONE: A SNAPSHOT IN TIME**

Stars brighten throughout their evolution (see, e.g., Hoyle 1958), with large luminosity changes outside of the MS phase. Therefore, the orbital distance that corresponds to the HZ, which is based on stellar irradiance and the host star's SED, moves outward with increasing stellar luminosity during a star's evolution (see, e.g., Villaver & Livio 2007, Danchi & Lopez 2013, Rushby et al. 2013, Luger & Barnes 2015, Ramirez & Kaltenegger 2014, 2016).

The observation-based empirical HZ is a very useful limit for preliminary identification of habitable exoplanets (discussed above). During the MS phase of the Sun, the orbital distance of the empirical HZ moves outward only by approximately 30%. Both Venus and Earth were in the HZ when the Sun started its MS phase. Venus is no longer in the HZ. The luminosity changes and corresponding HZ orbital changes can be much larger during the pre- and post-MS phases. **Figure 4** shows the changes in orbital distance of the empirical HZ for the pre-MS, MS, and post-MS HZ as well as one 3D model limit as dashed lines (Leconte et al. 2013a) for comparison. The stellar irradiance values of the 3D model HZ limits were used to derive those limits for the pre and post-MS. The times given in this section assume solar metallicity of the host stars. Stars with higher metallicity evolve more slowly, and therefore the time that a planet at a certain distance remains in the HZ is longer than for a star of lower metallicity (e.g., Danchi & Lopez 2013).

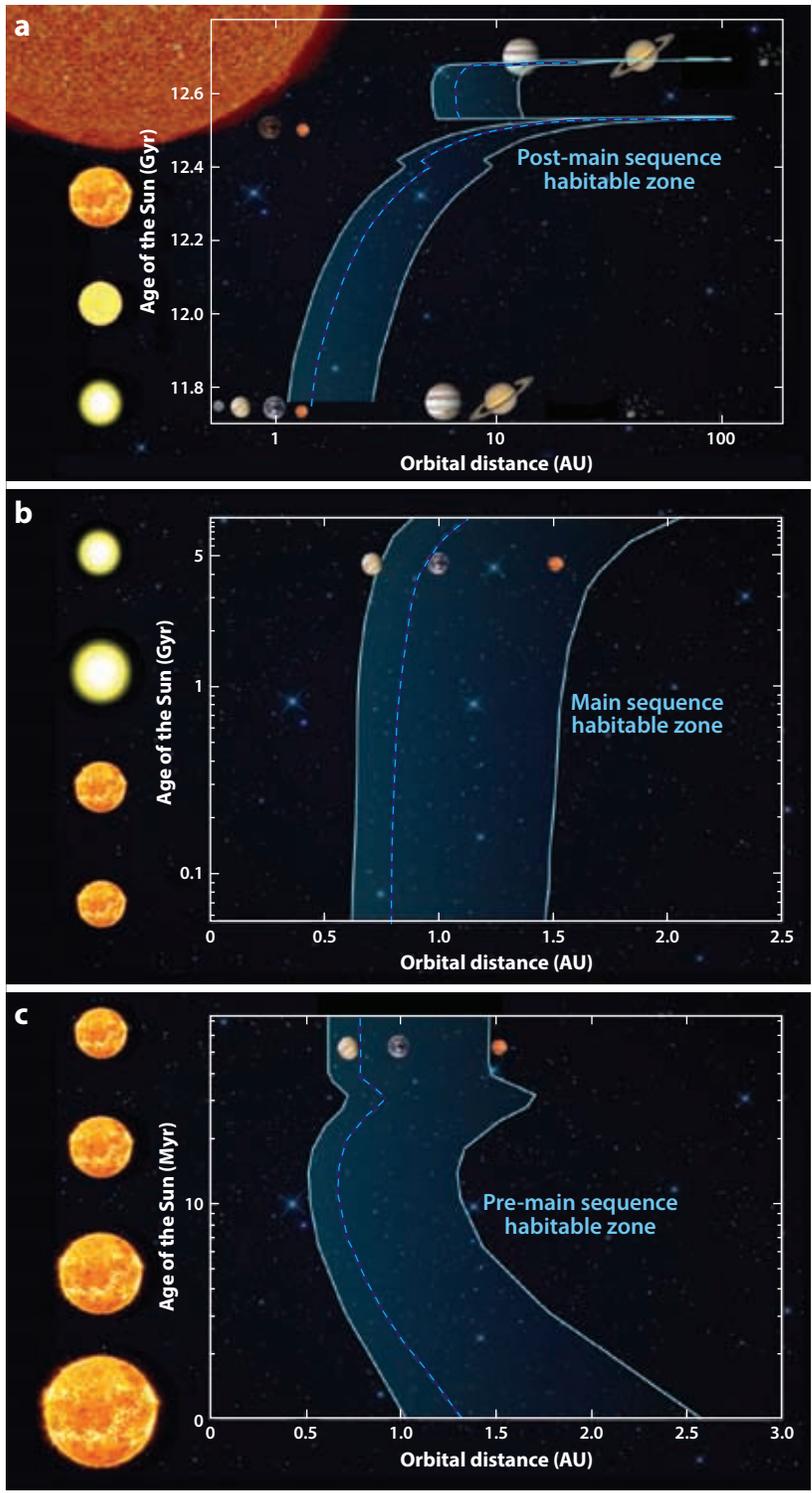

**Figure 4** Boundaries of the HZ for our Sun throughout its evolution: (*a*) post- MS, (*b*) MS, and (*c*) pre-MS. The orbital distances corresponding to the boundaries of the HZ evolve due to the star's changing luminosity and stellar energy distribution with age. Abbreviations: HZ, habitable zone; MS, main sequence. Figure courtesy of R. Ramirez.

Before a star arrives at the MS, its initial luminosity is higher, and thus its pre-MS HZ orbital distances are initially larger. Young exoplanets in the pre-MS HZ around cool stars would be easier to resolve with next-generation telescopes than planets in the MS of the same star type and could provide interesting insight into the planetary environment at very young ages. For example, the pre-MS HZ angular resolution for the about 40-Myr-old AP Col (M4.5, 8.4 pc away; Riedel et al. 2011) is approximately 8–24 milli-arcseconds (mAs) (i.e., Giant Magellan Telescope: 10 mAs, 1 μm; Thirty Meter Telescope and European-Extremely Large Telescope (E-ELT): <10 mAs, 1.3 μm; e.g., Lloyd-Hart et al. 2006, Kasper et al. 2008, Wright et al. 2010). The pre-MS orbital distance limits evolve due to the star's changing luminosity and SED: For the Sun, the inner edge of the HZ changes from 1 AU at the beginning to 0.6 AU. The outer edge correspondingly moves from 2.6 to 1.5 AU. The Sun's pre-MS stage lasted approximately 50 Myr. However, the pre-MS period of stellar evolution can last up to 2.5 Gyr for cool M stars, whereas accretion of planets in the HZ could only last several million years. This could potentially provide habitable conditions for up to 2 Gyr for cool M stars during the pre-MS phase (Ramirez & Kaltenegger 2014) before the star enters its MS phase.

For a host star of stellar type K5 and cooler, planets that are later located in the MS HZ receive stellar incident fluxes during the star's initial pre-MS phase that exceed the runaway greenhouse threshold, and thus may lose a substantial part of the water initially delivered to them (Ramirez & Kaltenegger 2014, Luger & Barnes 2015, Tian & Ida 2015). The incident stellar fluxes during the pre-MS phase of M stars are high enough to trigger runaway greenhouse conditions beyond the orbital distances of the MS HZ outer edge, which indicates that M-star planets in the MS HZ need to initially accrete more water than Earth did or, alternatively, have additional water delivered later to remain habitable. Triggering a runaway greenhouse effect does not necessarily mean that a planet has become uninhabitable, because water vapor could recondense when the stellar luminosity decreases. Whether water is permanently lost depends on how much hydrogen escapes from the top of the atmosphere, the basis of another lively debate in the literature (e.g., reviewed in Tian 2015).

The discovery of sub-Earth-sized planets around an 11-Gyr-old K-type MS star (Campante et al. 2015) illustrates that planets have formed early in the history of the Universe. Hotter star types, like the Sun, would already have moved onto the giant star branch of the post-MS at that age. Although none of the five planets found orbiting Kepler-444 are located in the HZ, this discovery raised the interesting question of where the HZ is located during the later stages of a star's evolution and whether these planets are detectable. Once a star leaves the MS and becomes a red giant, therefore, its luminosity increases and its HZ moves outward (**Figure 4**). Our Sun's post-MS HZ moves outward beyond the Kuiper belt (see also Lorenz et al. 1997, Stern 2003) to the outer regions of the Solar System. The post-MS HZ orbital distances are within the detection capabilities of direct imaging techniques, but the increased contrast ratio makes these planets challenging to detect (Ramirez & Kaltenegger 2016).

Assuming solar metallicity, a planet orbiting a post-MS solar analog can reside in the post-MS HZ for approximately 700 Myr, which is comparable to the amount of time for life to evolve on Earth (e.g., Mojzsis et al. 1996). For an M1 star, a planet could reside in the post-MS HZ for up to 9 Gyr, making planets outside the MS HZ of cool M stars interesting candidates for habitable conditions in the far future of the Universe.

As discussed above, the HZ is not the orbital distance at which life can evolve but at which life can be detected remotely if it exhibits surface or atmospheric signatures. Life might initially evolve in the subsurface on planets outside the MS HZ and be warmed during the post-MS phase, potentially uncovering hidden ecosystems, which could result in atmospheric biosignatures that can be detected remotely.

### 2.4. Detected Exoplanets in the Habitable Zone

When an exoplanet is detected, the stellar irradiance at its orbit can be calculated and compared to the incident stellar flux $S_{eff}$ that corresponds to the pre-MS, MS, and post-MS HZ limits. Equation 2 gives a third-order polynomial curve fit of the modeling results for A- to M-type host stars (Kopparapu et al. 2014, Ramirez & Kaltenegger 2014):

$$S_{eff} = S_{sun} + a \cdot T^* + b \cdot T^{*2} + c \cdot T^{*3} \qquad (2)$$

where T* = (T_eff − 5780) and S_sun is the stellar incident values at the HZ boundaries in our Solar System. **Table 2** shows the constants to derive the stellar flux at the HZ limits valid for $T_{eff}$ between 2,600 and 10,000 K: The inner boundaries of the empirical HZ (recent Venus, or RV) as well as an alternative inner edge limit for 3D GCMs (Leconte et al. 2013a) and the outer limits (early Mars). The outer HZ limit agrees in the 3D and 1D models and these entries are therefore not given in separate columns in **Table 2**.

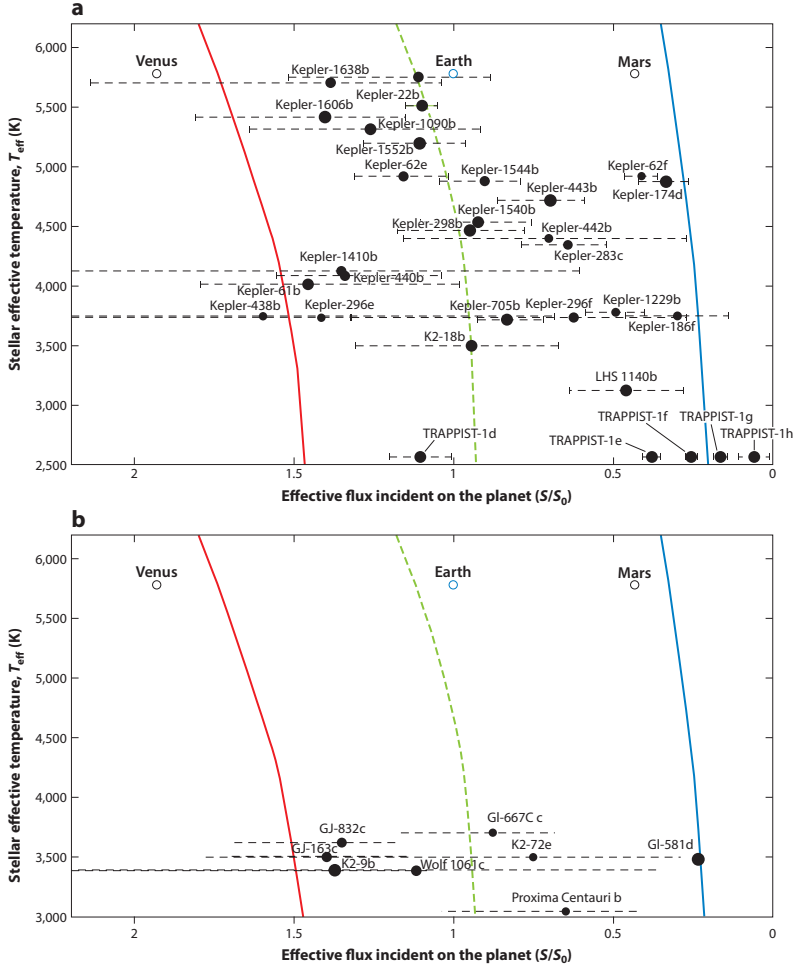

Figure 5 Detected exoplanets orbiting in the empirical habitable zone of their host stars (*solid red* and *blue lines*), as well as a three-dimensional model inner habitable zone limit (*dashed line*). (*a*) Transiting exoplanets with radii of 0–1, 1–1.5, and 1.5–2 $R_\oplus$ and (*b*) radial velocity exoplanets with minimum masses of 0–5 and 5–10 $M_\oplus$ (shown with differently sized dots for different mass and radius ranges). Data are from https://www.exoplanets.eu (accessed February 2017). Figure courtesy of R. Ramirez.

The orbital distance of the HZ boundaries can be calculated from $S_{eff}$ using Equation 3:

$$d = \sqrt{\frac{L/L_{sun}}{S_{eff}}} \qquad (3)$$

where $L/L_{sun}$ is the stellar luminosity in solar units and *d* is the orbital distance in astronomical units (AU).

**Figure 5** shows all detected planets in the HZ as a function of the $T_{eff}$ of their host star and the incident stellar irradiation $S_{eff}$; other recent lists of detected exoplanets can be found in the literature (e.g., Rowe et al. 2014, Kane et al. 2016). The error bars in **Figure 5** are derived from the uncertainties in stellar and orbital parameters given in the database and show the change in incident stellar flux estimates on the planet, depending on these uncertainties. Such measurement uncertainties influence the calculated incident stellar irradiance of exoplanets and whether a certain planet is considered to orbit within the star's HZ (see, e.g., Kaltenegger & Sasselov 2011, Von Braun et al. 2011, Boyajian et al. 2012, Dressing & Charbonneau 2013, Gaidos 2013).

The transiting HZ planet distribution (**Figure 5***a*) shows a very interesting diversity in both size and orbital parameters for exoplanets with radii below 2 $R_\oplus$. The three dot sizes shown in the figure correspond to 0–1, 1–1.5, and 1.5–2 $R_\oplus$. The RV HZ planet plot (**Figure 5***b*) is still sparsely populated, reflecting the difficulty of finding such light planets from the ground. Planets with minimum masses of 0–5 and 5–10 $M_\oplus$, respectively, are shown with differently sized dots.

An exciting aspect of the discoveries depicted in **Figure 5** is that they already show a fascinating, diverse set of planets in terms of mass, size, and orbital parameters, which indicates a wide variety of potentially habitable planetary environments.

## 2.5. How Many Potentially Habitable Planets Can We Expect?

One of the aims of exoplanet searches is to determine the frequency of rocky planets within the HZ, so-called $\eta_\oplus$. Several teams have published values for the frequency of planets in the HZ, with quite different results, as summarized in **Table 1**. Even though these teams used different data sets (because either some estimates were made with earlier data sets or some only used a subset of available data), the biggest difference in the numbers comes from the underlying assumptions made by different teams as to the stellar irradiance HZ limits and the values used to assert that a planet is rocky. All estimates indicate a large number of rocky exoplanets in the HZ, providing a large, diverse set of targets for characterization with next-generation telescopes (for details see **Table 1**).

# 3. OTHER WORLDS: CHANGES FROM THE INSIDE

Rocky exoplanets can be very different from Earth, for example, ocean-covered or desert worlds and planets with hydrogen-rich atmospheres, different geochemical cycles, increased volcanism, or a lack of plate tectonics. Some of these characteristics can strongly influence the chemical makeup of an atmosphere (e.g., high volcanism versus no volcanism) through different sources and sinks for atmospheric gases. Some of these ideas have been explored in the literature (see details below). A key to identifying and exploring such different rocky worlds is to assess how their observable spectra differ. Any biosignature must be analyzed critically for potential geological sources under conditions different from those on Earth (see, e.g., Kasting et al. 2014).

## 3.1. Detecting Geological Activity on Exoplanets

Geologic activity is crucial for a planet's maintained surface habitability because such habitability depends on the recycling of atmospheric gases like $CO_2$ (see details in Section 2.1.1). Earth-mass and even more massive planets should have sufficient radiogenic heating to maintain plate tectonics over long time periods (see, e.g., review by Korenaga 2013). There is a heated debate in the literature as to whether or not rocky planets more massive than Earth, super-Earths, could support plate tectonics. The mass of a planet influences its capability for tectonics in different ways in different models, making plate tectonics more likely due to increased plate velocities (Valencia et al. 2007, Valencia & O'Connell 2009, Van Heck & Tackley 2011) or less likely due to increased fault strength under high gravity (O'Neill et al. 2007, Noack & Breuer 2014) for more massive planets. Published works on both sides of the argument demonstrate the difficulty that arises from extrapolating from Earth's interior models. Whether or not tectonics alone could maintain a $CO_2$ cycle similar to that on Earth is unclear.

Although volcanism is not a biosignature (e.g., Jupiter's moon Io is very volcanically active, but there is no evidence that life exists there), it is a potential way to assess the geological activity of a rocky planet within the HZ. To trace geological activity on exoplanets, one could look for short-lived stratospheric gases erupted from highly active explosive volcanoes (Kite et al. 2008, Kaltenegger et al. 2010). Volcanic gases like $SO_2$ that can last for at least 100 days in the stratosphere of an Earth-like planet located in the HZ could generate temporary SO2 absorption features that are observable in transmission and emergent flux spectra for planets orbiting close by stars. Whether such stratospheric SO2 features could be detected also depends on whether or not atmospheric hazes evolved (Hu et al. 2013). Another approach proposed to detect volcanism in planetary atmospheres is to look for transient hazes generated by volcanism (Misra et al. 2015). However, hazes are not unique to volcanism and one would need to assess all false positives to determine whether such hazes had a volcanic cause.

Another open question is whether a planet can maintain a magnetic field that shields its atmosphere from erosion. How strong a planetary magnetic field needs to be to initially protect a planet's atmosphere from early stellar winds is another interesting unknown (see, e.g., Grießmeier et al. 2009, Lammer et al. 2011, reviewed in Tian et al. 2014). To generate a magnetic field using the iron core of a terrestrial planet, a minimum rotation speed and high heat flux at the core–mantle boundary are necessary (e.g., Stevenson 2003). For rocky planets more massive than Earth, a process like plate tectonics might be needed to cool a planet's surface sufficiently to maintain such high heat flux with high interior pressures limiting the development of a magnetic field (Elkins-Tanton & Seager 2008, Barnes et al. 2009, Stamenkovic et al. 2012). Observations of highly irradiated exoplanet atmospheres will be needed to constrain such model parameters, with initial concentrations and atmospheric loss rate due to stellar winds with and without magnetic fields being highly degenerate, except for extreme cases.

### 3.2. WATERWORLDS

Waterworlds in the HZ are completely novel objects that do not have a clear analog in our own Solar System. Nevertheless, the possible existence of Earth- and super-Earth-sized planets completely engulfed by a water envelope has long fascinated scientists and the general public alike (see, e.g., Kuchner 2003, Leger et al. 2004, Selsis et al. 2007, Zeng & Sasselov 2013) and exoplanet discoveries show densities suggesting a potentially large water inventory (see, e.g., Borucki et al. 2013, Kaltenegger et al. 2013). Although small Solar System bodies like Europa and Enceladus are composed of substantial quantities of water, none of them are in the HZ.

Water planets of Earth to super-Earth sizes in the HZ fall into at least two types of interior geophysical properties. In Type 1, the core–mantle boundary connects silicates with high-pressure phases of water (e.g., Ice VI, VII); that is, the liquid ocean has a high-pressure ice bottom. In contrast, in Type 2, the liquid ocean has a rocky bottom, though no silicates emerge above the ocean at any time, which is essentially a rocky planet in terms of bulk composition. Both subtypes could possess a liquid ocean outer surface, a steam atmosphere, or a full cover of surface Ice I, depending on their surface temperature.

Abbot et al. (2012) discussed how a carbonate-silicate cycle could be maintained with up to 95% of a planet's surface covered by water, suggesting by extrapolation a possible $CO_2$ cycling for Type 2 waterworlds via the rocky ocean floor. Such planet models show a varying timescale for a carbon cycle, depending on their initial carbon inventory and temperature (Foley 2015). However, for Type 1 waterworlds, that is, super-Earths with water mass fractions like those of Earth or higher, the deep oceans are separated from the rocky interior via a layer of high-pressure ices. Therefore, an alternative mechanism to modulate abundant gases like $CO_2$ and $CH_4$ would be required in waterworlds. Such a mechanism would depend on the properties of their clathration in water over a range of very high pressures. Under very high pressures (above 0.6 GPa), common to super-Earth-sized water planets, clathrates undergo a phase transition in their structure to a form known as filled ice that can transport $CH_4$ and $CO_2$ (Levi et al. 2013, 2017).

This should enable effective cycling of $CO_2$ through the atmosphere and oceans on water planets residing in the HZ (e.g., with water oceans at the surface) and thus provide an alternative atmospheric concentration feedback mechanism to regulate $CO_2$ on water planets. The timescales and effectiveness of such cycling is still an open question (Kitzmann et al. 2015, Levi et al. 2017). A lively debate in the literature addresses how the interiors of such worlds might work, regarding the ability to cycle gases from the rocky core to the atmosphere (Valencia et al. 2007, Grasset et al. 2009, Fu et al. 2010, Rogers & Seager 2010, Kaltenegger et al. 2013, Tackley et al. 2013, Levi et al. 2013, 2017) and whether such planets could be habitats (Abbot 2016, Abbot et al. 2012, Kaltenegger et al. 2013, Noack et al. 2016). First analyses of what atmospheric spectra of waterworlds might look like and whether they could be distinguished from land-dominated worlds even in transit, where surface albedo would not be detectable, depend strongly on whether long-term $CO_2$ cycling can be maintained (Kaltenegger et al. 2013).

### 3.3. EXOMOONS AS REMOTELY DETECTABLE HABITATS

The characterization of extrasolar giant planets (EGPs) has revealed a fascinating, diverse set of planets (e.g., reviewed in Burrows 2014, Crossfield 2015). EGPs might have exomoon satellites. Such exomoons have not been detected yet but could be found through a variation of exoplanet transit timing and/or duration due to the orbit of the planet around the planet–moon barycenter (Sartoretti & Schneider 1999, Agol et al. 2005, Holman & Murray 2005, Kipping 2009), light curve distortions (Szabo et al. 2006), planet–moon eclipses (Cabrera & Schneider 2007), microlensing (Han 2008), pulsar timing (Lewis et al. 2008), distortions of the Rossiter–McLaughlin effect of a transiting planet (Simon et al. 2009), and potentially direct detection of exomoons at specific wavelengths (Agol et al. 2015). A detailed study (Kipping et al. 2009, 2012) using transit time duration measurements found that exomoons down to 0.2 $M_\oplus$, double the size of Mars, may be detectable around EGPs in the HZ of their host star with the Kepler mission or equivalent photometry.

Such exomoons could orbit in the circumstellar HZ (e.g., Reynolds et al. 1987, Williams & Kasting 1997, Kaltenegger 2000, Scharf 2006, Kaltenegger & Haghighipour 2013, Forgan & Kipping 2013, Hinkel & Kane 2013, Heller et al. 2014, Heller & Pudritz 2015). A habitable moon would need to retain its volatiles. This depends on its mass, the charged particle flux it receives, whether it maintains a magnetosphere, and its position with respect to the planet's magnetosphere, among other factors; these factors lead to a lower mass limit between 0.12 and 0.23 $M_\oplus$ (e.g., Williams & Kasting 1997), above which moons can potentially be Earth analog environments, which we will call Earth-like moons here. Several groups have shown the long-term dynamic stability of hypothetical satellites and moons up to Earth-mass orbiting EGPs (e.g., Barnes et al. 2002, Dvorak et al. 2010) as well as discussion on a potential influence of tidal heating on such moons (e.g., reviewed in Henning et al. 2009, Henning & Hurford 2014, Chen et al. 2014, Heller et al. 2014).

Close-by small stars are interesting candidates for detection and subsequent characterization of potentially habitable exomoons. As for exoplanets, cool stars are interesting targets to find and characterize potentially habitable exomoons because of the small orbital distance of their HZs, which increases the transit probability as well as the transit frequency per observation time and the increased contrast ratio due to the small size of the star. In addition, even if HZ planets for cool stars were synchronously locked to their host star, exomoons orbiting them in turn would be synchronously locked to their planet, not the host star, illuminating the whole moon as compared to only partial illumination of a synchronous planet. Note, however, that synchronous rotation may not be a problem for habitability on a planet that has a substantial atmosphere, like Earth, that distributes heat evenly on its surface (see, e.g., Joshi 2003, Scalo et al. 2007, Kite et al. 2009, Edson et al. 2012).

An interesting question is whether one could disentangle the moon's spectrum from that of its parent planet if they are seen together. Generally, an unresolved moon will influence the spectrum of its planet, as the spectra of both will appear to come from the planet. Even though the features of an EGP and an Earth-like exomoon are expected to be different (e.g., Williams & Gaidos 2008, Robinson et al. 2011), the two-order magnitude difference in feature strength would make it extremely difficult to detect a moon's atmospheric spectral feature embedded in an EGP planet's spectrum if the EGP spectrum is not fully understood. Highly reflective moons and EGPs with low albedos can improve the planet–moon contrast ratio and could make detection possible if temporal intensity and spectral information were available (see, e.g., Moskovitz et al. 2009, Agol et al. 2015).

However, if a habitable moon orbits its planet at a distance that allows for spatially separate transit events, transmission spectroscopy will be a unique potential tool to screen exomoons for habitability in the near future (Kaltenegger 2010). If the moon can be spatially resolved, characterizing an Earth-sized habitable moon would provide challenges comparable to characterizing an Earth-sized planet.

Viewing the existence of moons as important to stabilize the obliquity of a planet (Ward 1982) might be an overly Earth-centric outlook. Lissauer et al. (2012) found that without the presence of a moon, obliquity variations are still significantly constrained. In addition, subsurface conditions, for example, of an ocean, should not be influenced severely by changes in a planet's obliquity, and life could alternatively evolve for changing surface conditions.

## 4. CHARACTERIZING A HABITABLE WORLD

The following steps can be undertaken to set the planetary atmosphere in context. After an exoplanet's detection, one can first calculate the incident stellar irradiance the planet receives that puts it in- or outside the HZ limits as discussed in Section 2.2.

First, with regard to reading exoplanet colors, low-resolution spectra of exoplanets are likely the first post detection measurements. Color-color diagrams (three or four channels) are used to classify many astronomical objects and can be used to initially compare the color of extrasolar planets to Solar System bodies (see, e.g., Traub 2003, Lundock et al. 2009, Fujii et al. 2011, Hegde & Kaltenegger 2013, Krissansen-Totton et al. 2016, Madden & Kaltenegger 2017, O'Malley-James & Kaltenegger 2017a). Planets in the Solar System can be grouped by their color spectra into different categories; for example, ice giants would occupy a different part of such a diagram than rocky bodies. Whether such an initial comparison of exoplanet spectra to Solar System planet spectra would also allow the inference of exoplanet properties is an open question, due to our limited sample of characterized planets for now. However, we could use it for an initial comparison, which would become more insightful when the color-color diagram could be populated with characterized exoplanets.

Second, orbital flux variations can distinguish bodies with an atmosphere from airless ones. Orbital phase curves have already been used to explore the atmosphere of brown dwarfs and EGPs (e.g., reviewed in Knutson et al. 2007, Crossfield 2015). A next step will be to use them to characterize smaller, rocky planets. If a planet's thermal radiation is observed, flux measurements over a planet's orbit can distinguish planets with atmospheres from those without by the amplitude of the observed orbital variations (**Figure 6**). The amount of reflected light depends on the lit area and averaged reflectivity of the planet. Phase-related variations in a planet's or planet–moon system's visible flux are therefore in phase with the amount of lit surface. The thermal radiation of the planet and moon depends on its overall area as well as the averaged hemispheric temperature. Therefore, strong variations of the IR thermal flux with the phase reflect strong temperature differences in the day and night hemispheric temperatures, a consequence of, for example, the absence of a dense atmosphere or ocean (e.g., Gaidos & Williams 2004, Selsis 2004).

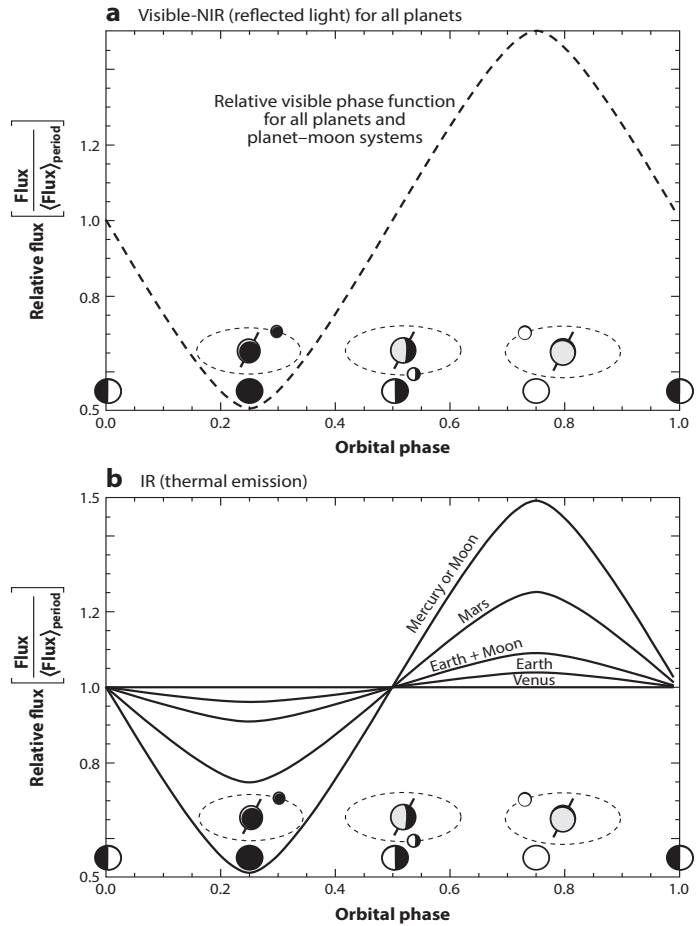

**Figure 6** Schematics of relative orbital light curves for planets and planets with unresolved moons in a circular orbit with and without an atmosphere in (*a*) reflected light and (*b*) thermal emission, assuming blackbodies. Unresolved moons can also influence the detected light curve. Data from Selsis et al. (2008). Abbreviations: IR, infrared; NIR, near-infrared.

The light curve of Earth and Venus in our Solar System would exhibit nearly no measurable phase-related variations of their thermal emission. However, Mars shows intensity fluctuation in phase with the lit surface, indicating a thin atmosphere. As shown in **Figure 6**, the combined flux from the Moon and Earth shows some intensity fluctuations due to the unresolved Moon. Unresolved moons in general can influence the phase curve, but how strongly depends on the characteristics of the planet and moon(s), which can lead to confusion for very large moons with surfaces that have very low heat capacity (e.g., Moskovitz et al. 2009). However, such contributions to the combined IR phase curve by the lit part of a moon vary with orbital phase and thus could be identified. Such changes might even be used to detect exomoons (see, e.g., Agol et al. 2015). Orbital phase curves also put constraints on other planetary characteristics like its obliquity (see, e.g., Gaidos 2004, Selsis et al. 2011, Schwartz et al. 2016).

Finally, with regard to deriving the radius and temperature of an exoplanet, for transiting planets, the radius of the planet is known as a fraction of its host star's radius. The radius of non-transiting planets can be estimated using a low-resolution thermal spectrum. Using a blackbody approximation, a thermal spectrum will constrain the temperature of the emitting layer, and therefore the emitting area and planetary radius (see **Figure 7**). The accuracy of these constraints will depend on the quality of the fit (and thus on the sensitivity and resolution of the spectrum) and whether the emitting layer is the planet's surface.

For transparent atmospheres like those of Earth and Mars, the surface temperature can be observed in the IR (see **Figure 8**). This can also be done for exoplanets if spectral windows can be identified that probe the surface. Venus appears cooler than Earth because the emitting layer is in its atmosphere, not the surface. For present-day Earth, the IR flux between 8 and 11 μm probes the surface temperature. This window would, however, become opaque at high $H_2O$ partial pressure (e.g., the inner part of the HZ, where a lot of water is vaporized) and at high $CO_2$ pressure (e.g., a very young Earth or the outer part of the HZ), closing our ability to assess the surface temperature. If no such windows exist or the thermal radiation cannot be measured, one can only use models, corresponding to the atmospheric composition of a planet, to estimate surface temperatures. Alternatively, one can constrain the radius very loosely using a mass range derived from the minimum mass measurements and assuming a certain composition of the planet.

A higher-resolution spectrum can be used to identify the compounds of the planetary atmosphere. In that context, we can then test if we have an abiotic explanation of all compounds seen in the atmosphere of such a planet. If we do not, we can work with the exciting biotic hypothesis.

### 4.1. EARTH SEEN AS A PALE BLUE DOT

In the emergent flux, we observe the starlight reflected off the planet in the UV to NIR and the planet's emitted flux in the IR, dependent on the planet's temperature (**Figures 7–9**). Emergent flux combines the radiation from up to a full hemisphere and probes different atmospheric depths, including the surface, unlike a transmission spectrum. In transmission, we see the starlight filtered through the planet's atmosphere. Transmission spectra predominantly probe the upper atmosphere of planets, depending on the system's geometry and planet's atmospheric density

Reflectance spectra mix stellar and planetary lines because a planet's visible flux is reflected starlight. Therefore, plots that only show relative reflectivity in the visible are informative but can be misleading because they do not show how much light the host star emits at different wavelengths. However, such plots can easily be multiplied by the host star's SED to get the complete picture. **Figure 7** gives the absolute flux for a Solar System analog at 10 pc. The Sun's flux is several orders of magnitude higher in the visible than the IR. Therefore, the absolute flux of Earth is higher in the visible than in the IR; however, the contrast ratio between Earth and the Sun improves in the IR over the visible by a factor of approximately 1,000.

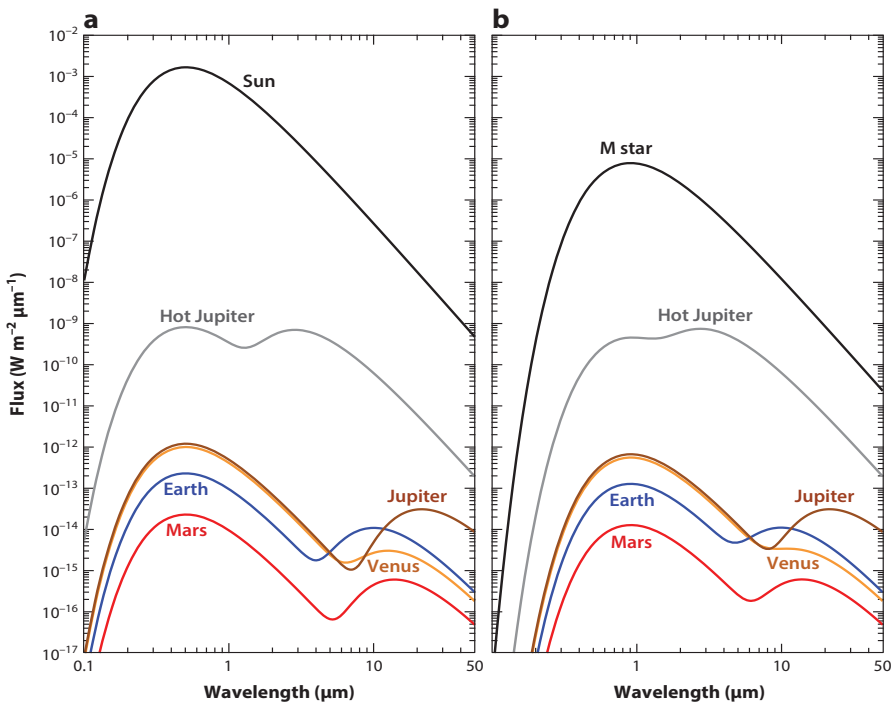

**Figure 7** Absolute flux comparison of Jupiter, Venus, Earth, and Mars in our Solar System as well as a hot extrasolar giant planet shown here as blackbodies for the (*a*) Sun and (*b*) M0 host star at a distance of 10 pc on a log-log scale, assuming constant reflectivity and temperature. Panel *b* shows the decrease in planet–star contrast ratio for planets orbiting cool stars.

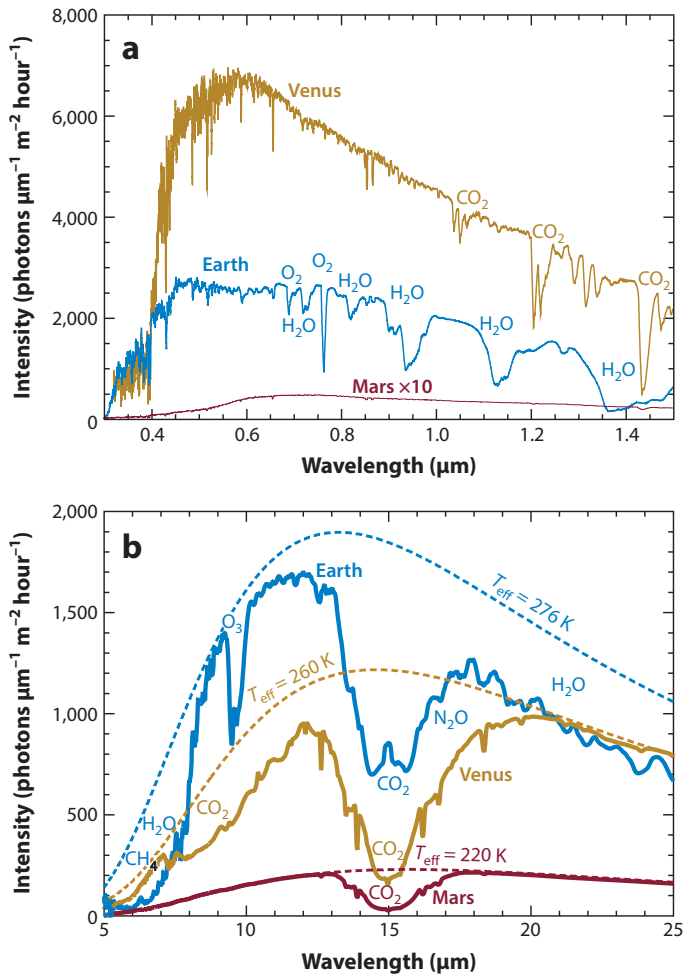

**Figure 8** Spectra of Earth, Venus, and Mars at a resolution ($\lambda/\Delta\lambda$) of approximately 100 in the (*a*) visible to near infrared bands (the reflected flux of Mars has been multiplied by 10 to appear) and (*b*) thermal emission spectrum of the planets and blackbody emission of a planet of the same radius at the maximum brightness temperature of the spectrum. Data from Selsis et al. (2008).

The Earth–Sun contrast ratio in emergent flux is approximately $10^{-7}$ in the thermal IR (~10 μm), and approximately $10^{-10}$ in the visible (~0.5 μm) (**Figure 7**). Comparing the absolute flux of Earth, Mars, Venus, and Jupiter in our Solar System and adding an EGP model for a Jupiter-sized planet with a 1,500-K surface temperature and an albedo of 0.3 for simplicity, **Figure 7** shows how the flux as well as planet–to–host star contrast changes for different types of exoplanets as a function of wavelength. Keeping the planets at the same incident stellar flux as that in our Solar System, **Figure 7** shows how the planet–to–star flux ratio changes for a cool M0 host star. The Earth–M0 star contrast ratio is approximately $10^{-5}$ in the thermal IR (~10 μm) and approximately $10^{-8}$ in the visible (~0.5 μm). The Earth–F0 star contrast ratio is approximately $10^{-8}$ in the thermal IR (~10 μm) and approximately $10^{-11}$ in the visible (~0.5 μm).

For super- Earths, with 2 $R_\oplus$, the contrast ratio decreases by a factor of 4 compared to Earth-sized planets. For smaller planets, the contrast ratio increases owing to their smaller surface area. The HZ for cooler stars is at smaller and for hotter stars at larger orbital distances, which increases the transit frequency of HZ planets for cooler stars but also increases the resolution required to detect them individually.

**Figure 8** shows the visible and IR emergent spectra of Venus, Earth, and Mars models for a resolution of approximately 100 for comparison (see also Selsis et al. 2008). Different wavelengths have different sensitivity to clouds, Rayleigh scattering, or hazes and show different atmospheric chemicals, as discussed in detail below. The abundance of a chemical needed to detect a spectral feature at a certain resolution varies (see, e.g., Des Marais et al. 2002, Selsis et al. 2002, Kaltenegger et al. 2007). The observable depth of spectral features in reflected light is dependent on the abundance of a chemical as well as the incoming stellar radiation at that wavelength. In thermal emission, the depth of spectral features depends on the abundance of a chemical as well as the temperature difference between the emitting/absorbing layer and the continuum.

Properties like the temperature of the emitting surface can be probed in the IR. Surface reflection can be detected in the visible for transparent atmospheres. The trade-off between contrast ratio, detectable spectral features, and design in the different wavelengths is not discussed here but leads to several different configurations for space-based mission concepts.

For Mars and Venus, only $CO_2$ features are observable at a resolution of 100, whereas Earth shows absorption features of $O_2$ and $H_2O$ in the visible and $H_2O$, $CH_4$, $O_3$, and $CO_2$ in the IR (**Figure 8**).

Earth's emergent spectrum from the visible to IR is shown in **Figure 9** and can be observed when Earth is seen as an individual planet in direct imaging or close to a secondary eclipse. The model (see Kaltenegger et al. 2007) is compared to observations for visible (Woolf et al. 2002) and NIR earthshine (Turnbull et al. 2006) and a space-based emission IR spectrum (Christensen & Pearl 1997), respectively. Other measurements of Earth's emergent spectra are similar to the ones shown in **Figure 9** (see, e.g., Arnold 2008, Robinson et al. 2014, Palle et al. 2016).

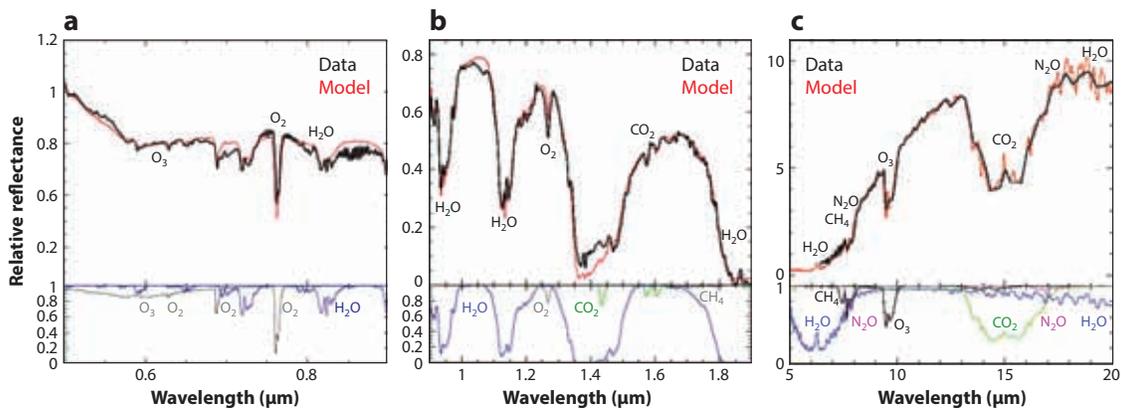

**Figure 9** (*a*) Visible, (*b*) near-infrared reflectivity, and (*c*) infrared emission spectra of Earth (model shown in *red*). Data (*black*) are from earthshine measurements and space measurements, respectively (data from Kaltenegger et al. 2007).

### 4.2. EARTH SEEN AS A TRANSITING PLANET

Earth seen as a transiting planet would dim the Sun's light for a maximum of 12.6 hours and produce a maximum transit depth of 0.0084%, which corresponds to the area ratio of the Sun to Earth. For cool stars, the transit depth increases to approximately 0.022% for an M0V and to 1.3% for an M9V host star, but the maximum transit duration decreases to 5.4 and 0.4 hours, respectively, whereas the transit frequency increases.

A transiting planet appears bigger at the wavelengths at which molecules in its atmosphere absorb light. The effective atmospheric thickness translates into an effective detected planetary radius, $R_{eff}$, expressed in Earth radii, $R_p$ (**Figure 10**). For present-day Earth, we find a good agreement between 1D models using Earth's average atmospheric composition and temperature structure (1979 Earth model) and available data for both emergent and transmission spectra.

Earth's transmission spectrum from the UV to IR (**Figure 10**) shows how Earth would look when observed as a transiting planet. In transmission, the apparent size of the Earth depends strongly on the wavelength of observation. In the far-UV, at approximately 150 nm, $O_2$ absorption increases the effective planetary radius of a transiting Earth by approximately 180 km (Betremieux &Kaltenegger 2014) versus 27 km in the visible at 760 nm (Kaltenegger & Traub 2009, Betremieux &Kaltenegger 2014), 14km in the NIR, and between 50 and 12 km in the IR due to absorption and refraction (Kaltenegger & Traub 2009). This translates to a 2.6% change in effective planetary radius over the UV–IR. For Earth, the $O_2$ absorption dominates the UV absorption shortward of 200 nm. For planets without $O_2$, the far-UV would still show strong absorption by other molecules, such as $CO_2$ and $H_2O$, increasing the planetary radius significantly in the UV. However, as discussed above, stellar incident flux sets the detectable absolute flux for a planet's transiting as well as reflected spectra flux. Less stellar flux is emitted by a cooler star in short wavelength range and the UV, except for active stars, making the increase in Earth's radius less effective for detection in the UV wavelength range. In addition the visible to IR is a more easily accessible wavelength range for remote observations.

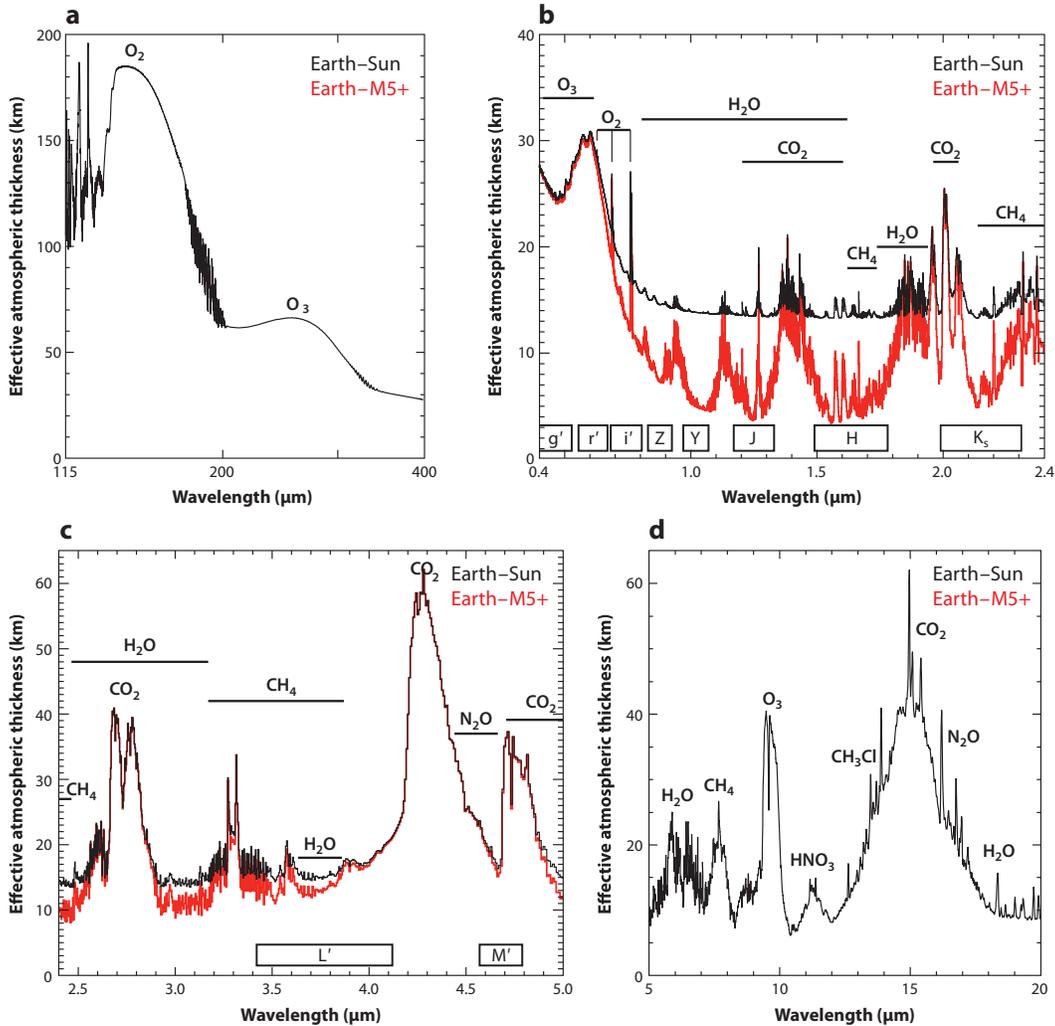

**Figure 10** (*a*) Ultraviolet, (*b*) visible, (*c*) near-infrared, and (*d*) infrared transmission spectra of a transiting Earth analog orbiting the Sun (*black line*). In the visible to infrared spectrum (*b,c*), refraction limits the depth the atmosphere of an Earth-analog planet can be probed down to significantly (the *red line* shows a model that does not take refraction into account for the Sun–Earth case). Spectral coverage to the 50% transmission limit of several commonly used filters is indicated by the black rectangles. Due to the geometry of the star–planet size and separation, refraction does not limit the depth of the atmosphere transiting Earth-analog planets can be probed to for planets orbiting a cool M-type host star (M5 and cooler) (*red line*). (a) (Betrimeux & Kaltenegger 2013), (b) and (c) (Betremieux & Kaltenegger 2014), (d) (Kaltenegger & Traub 2009).

No data for a transmission spectrum of Earth as it would be seen by a remote observer exists; therefore, the models need to be validated by data that can be added to mimic a transiting Earth. Data taken by ATMOS 3 (Atmospheric Trace Molecule Spectroscopy 3) on a shuttle mission can verify the transmission geometry spectra using the individual observations of the Sun observed through different heights of Earth's atmosphere. Summing these observations up, one can mimic the total transmission spectrum of Earth (see details in Kaltenegger & Traub 2009).

Note that an earthshine data set taken to mimic Earth as seen in transit (Palle et al. 2009) is misrepresenting what the transmission spectrum of Earth would look like to a distant observer, because light rays that hit the Moon and get reflected back to Earth must be bent strongly in Earth's atmosphere (see e.g. Garcıa Munoz et al. 2012, Betremieux & Kaltenegger 2013, Misra et al. 2014). Such rays therefore predominantly represent the dense part of Earth's atmosphere. For a transiting planet, light rays that are bent so strongly will not reach a distant observer (ibid).

**4.2.1 How Deep Can One Probe Earth's Atmosphere in Transmission?** Measurements of transiting hot EGPs already offer the possibility of characterizing their upper atmosphere (reviewed in, e.g., Burrows et al. 2014, Crossfield 2015). Such transit observations should be able to access biosignatures in the atmospheres of Earth-like bodies (e.g., Ehrenreich et al. 2006, Kaltenegger & Traub 2009, Palle et al. 2009, Vidal-Madjar et al. 2010, Rauer et al. 2011, Garcıa Munoz et al. 2012, Hedelt et al. 2013, Snellen et al. 2013, Betremieux & Kaltenegger 2013, 2014, Misra et al. 2014, Rodler & Lopez-Morales 2014).

As light rays traverse an atmosphere, they are bent by refraction from the major gaseous species, which to first order are proportional to the density of the gas so that the deepest atmospheric regions bend the rays the strongest. Refraction and absorption limit how deeply one can probe a transiting habitable planet. In transit geometry, the star is an extended source with respect to the observed planetary atmosphere and the observer is infinitely far away. During transit, refraction from the deeper atmospheric regions deflects light away from the observer (e.g. Garcıa Munoz et al. 2012, Betremieux & Kaltenegger 2014, Misra et al. 2014). This effect masks molecular absorption features that originate below a certain altitude, depending on the density and composition of the atmosphere and the geometry of the planet–star system. The lowest 12 km of Earth's atmosphere are not accessible to a distant observer when Earth transits the Sun, because no rays below that altitude can reach an observer from an Earth–Sun geometry.

The angular size of the host star with respect to the exoplanet determines the critical deflection by which a ray can be bent and still reach the observer as parallel light rays (see Garcıa Munoz et al. 2012, Betremieux & Kaltenegger 2014, Misra et al. 2014) and varies depending on the host star for planets in the HZ. In a first-order approximation, the height that can be probed scales with approximately $1/T_{\rm eff}^2$ for exoplanets with similar incident stellar flux at the top of their atmospheres (Betremieux & Kaltenegger 2014). Thus, observers can more deeply probe the atmosphere of similar planets orbiting cooler stars for a given atmospheric profile (red line in **Figure 10**).

Water will be one of the hardest features to find in a transiting spectrum from an Earth–Sun analog geometry because it is concentrated in the lowest 10–15 km of Earth's atmosphere; hence, only the amount above 12 km can increase the planetary radius above the refraction threshold in transit. Clouds do not significantly affect the transmission spectra of Earth (to a distant observer), because they are located below the 12-km threshold.

# 5. BIOSIGNATURES: HOW TO DETECT SIGNATURES OF LIFE ON OTHER WORLDS

Signs of life that modify the atmosphere or the surface of a planet and thus can be remotely detectable are key to finding life on exoplanets or exomoons. Observations of our Earth with its diverse biota serves as a Rosetta stone to identify habitats. **Figures 9** and **10** show that some atmospheric species exhibiting noticeable spectral features in our planet's spectrum result directly or indirectly from biological activity, the main ones being $O_2$, $O_3$, $CH_4$, $N_2O$, and $CH_3Cl$ (see, e.g., Des Marais et al. 2002). Spectroscopy can reveal the presence of specific molecules and atoms by identifying their characteristic energy transitions.

Sagan et al. (1993) analyzed an emergent spectrum of Earth, taken by the Galileo probe, searching for signatures of life and concluded that the large amount of $O_2$ and simultaneous presence of $CH_4$ are strongly suggestive of biology, as Lovelock (1965) had suggested earlier. The concentrations of oxygen and methane are far from equilibrium in Earth's atmosphere. On short timescales, the two species react to produce carbon dioxide and water and therefore must be constantly replenished to be maintained at detectable concentrations.

The term biosignatures is used here to mean gases that are produced by life, accumulate in the atmosphere, are not readily mimicked by abiotic processes, and can be detected by space telescopes, for example, the $CH_4$ + $O_2$ (Lederberg 1965, Lovelock 1965) or $CH_4$ + $N_2O$ (Lippincott et al. 1967) pairs. It is their quantities and detection along with other atmospheric species in the planetary context that solidify a biological origin. Bioindicators is used here to mean atmospheric signatures that indicate habitability, like $CO_2$ and $CH_4$, which can be produced by life but also in large quantities by abiotic processes. The two gases, $H_2O$ and $CO_2$, are important both as greenhouse gases as well as potential sources for high $O_2$ concentrations through photolysis. These examples already show that the combination of our knowledge of the host star and the planetary environment will be crucial to understanding other planets, moons, and habitats.

### 5.1. ATMOSPHERIC BIOSIGNATURES: DETECTABLE GASES AS SIGNS OF LIFE

Due to the vast interstellar distances, our first exoplanet spectra of rocky planets in the HZ will most likely have low SNR. First-generation space mission observations will also have low resolution ($R < 100$). Such low-resolution spectra will be comparable to the early stages of Solar System planetary science, when we only had disk-integrated spectra from Solar System planets and moons and were confronted with a wide range of vastly different worlds that we had to learn to understand.

The situation is slightly different when trying to find biosignatures from the ground, because the challenge is to tell them apart from biosignatures in our own atmosphere. Ground-based telescopes like the ELTs will have to use high-resolution spectra to identify exoplanet biosignatures via their Doppler-shifted lines, which requires combining observations over long periods for Earth analog planets (e.g., Snellen et al. 2013, Rodler & Lopez-Morales 2014) to achieve the required SNR.

Life produces thousands of molecules on Earth (e.g., reviewed in Seager et al. 2016), but to detect them over interstellar distances, they must modify the atmosphere in a detectable way that allows us to distinguish them from purely abiotic processes.

Due to the wide possible range of characteristics of other worlds and SED and UV flux of their host stars, this question becomes even more complex. Atmospheric models explore which molecules could function as biosignatures over interstellar distances and whether other environments could produce false positives abiotically.

As discussed above, Lovelock (1965) suggested looking for a by-product of metabolism, the simultaneous presence of $O_2$ and a reduced gas like $CH_4$ in a planet's atmosphere, as strong evidence that the planet is inhabited. Lippincott et al. (1967) extended this to specific gas pairs like $O_2$ and $CH_4$ or $O_2$ and $N_2O$. These gases are many orders of magnitude out of equilibrium with each other and are all predominantly produced biologically. Reduced gases and oxygen have to be produced concurrently to be detectable in low resolution in the atmosphere of habitable exoplanets, because they react rapidly with each other and would not stay observable if not continuously produced (see detailed discussion below on generalizing this idea).

On present-day Earth, $O_2$ and $CH_4$ are mostly produced by organisms and can be detected with low-resolution spectra ($R < 100$) depending on the wavelength (see, e.g., Des Marais et al. 2002). $N_2O$ is not easily observable for Earth (see details below; e.g., Segura et al. 2005, Kaltenegger et al. 2007, Grenfell et al. 2011). It is crucial to differentiate this combination of gases from the individual gases alone. One of those gases by itself, for example, $O_2$, does not constitute a biosignature, because each individual gas can be produced abiotically and could build up in an atmosphere without life (see, e.g., Harman et al. 2015).

Secondary metabolic products such as methyl chloride ($CH_3Cl$; e.g., Segura et al. 2005b), dimethyl disulfide (Pilcher 2003), dimethyl sulfide, and other sulfur gases (Domagal-Goldman et al. 2011) have also been studied as potential biosignatures. However, these gases are produced in small amounts on Earth and photolyzed sufficiently rapidly that they are not expected to build up to observable concentrations in a planet's atmosphere (e.g., Kasting et al. 2014). The extreme-UV (EUV) and far-UV radiation drives atmospheric photochemistry and varies depending on host star type and age (see, e.g., France et al. 2013, Linsky et al. 2013), which could change the buildup of different chemicals and their detectability as discussed below (see, e.g., Segura et al. 2010).

Technology biosignatures are, for example, the manufactured gases, like CFCs ($CCl_2F_2$ and $CCl_3F$) in our present-day atmosphere, which are only produced by technology. Such chemicals could be used as markers of advanced civilizations. They have absorption features in the thermal IR waveband; however, their abundances are very low, and they are extremely hard to detect spectroscopically (see, e.g., Des Marais et al. 2002, Selsis et al. 2002, Kaltenegger et al. 2009).

Another very interesting approach is to consider all stable and potentially volatile molecules as potential biosignatures and examine thousands of molecules produced on Earth to assess whether any of those could be a new biosignature for different chemical environments (see Seager et al. 2016). A first step is to compile a list consisting of all molecules that are stable and potentially volatile, and a second is to identify those molecules produced by life on Earth. In further steps, the viability of these molecules as a biosignature gas on exoplanets will need to be assessed, and the strength and wavelength range of that biosignature gas's spectroscopic signature will need to be modeled. This thorough approach could find new atmospheric biosignatures for very different environments, which would then need to be tested for false positives.

### 5.2. WHERE TO SPOT BIOSIGNATURES

As for any spectral features, the amount of chemicals needed to show a feature varies depending on the wavelength. The depth of spectral features in reflected light (UV to NIR for Earth) is dependent on the abundance of a chemical as well as the incoming stellar radiation at that wavelength. In the IR, Earth's spectrum shows our planet's thermal emission; therefore, the depth of spectral features depends on the abundance of a chemical as well as the temperature difference between the emitting/absorbing layer and the continuum.

For an Earth-like biosphere, the main detectable atmospheric chemical signatures that in combination indicate habitability are $O_2/O_3$ in combination with $CH_4$ or $N_2O$ (see discussion in Section 5.1). For present-day Earth atmosphere models, detecting the combination of $O_2$ or $O_3$ in combination with $CH_4$ requires observations in the visible to NIR from 0.7 to 3 μm to include the 2.4 μm $CH_4$ feature or observations in the IR between 5 and 10 μm (**Figures 9** and **10**). In addition, $CH_3Cl$ and $N_2O$ show weak absorption features in the IR between 7 and 17 μm. A wider wavelength coverage would give context to such detected features, like the IR $CO_2$ feature at 15 μm.

Biosignatures and bioindicators on Earth can be detected in several different wavelengths (**Figures 9** and **10**): $H_2O$, $O_3$, and $O_2$ in the visible (400 nm–2 μm), $CH_4$ and $CO_2$ in the NIR (2–4 μm), and $CO_2$, $H_2O$, $O_3$, $CH_4$, $N_2O$, and $CH_3Cl$ in the thermal IR (4–20 μm).

The UV wavelength range is very sensitive to small molecular abundances; therefore, it is generally unreliable for biosignatures. As an example, under certain conditions, the UV $O_3$ feature could be detected in prebiotic Earth models (Domagal-Goldman et al. 2014), whereas the same models did not produce detectable abiotic visible $O_2$ or IR $O_3$ features.

In the visible wavelength range, from 400 nm to 2 μm, the strongest $O_2$ feature is the saturated Frauenhofer A band at 0.76 μm, with a weaker feature at 0.69 μm. $O_3$ has a broad feature, the Chappuis band, which appears as a shallow triangular dip in the middle of the visible spectrum from approximately 0.45 to 0.74 μm. Methane at present terrestrial abundance (1.65 ppm) has no significant visible absorption features, but at high abundance, it shows bands at 0.88 and 1.04 μm, detectable, for example, in early Earth models (**Figure 10**). In addition to biosignatures, $H_2O$ shows bands at 0.73, 0.82, 0.95, and 1.14 μm. $CO_2$ has negligible visible features at present abundance, but in a high- $CO_2$ atmosphere of 10% $CO_2$, as in early Earth evolution stages, the weak 1.06 μm band could become detectable.

In the NIR, from 2 to 4 μm, there are $CH_4$ features at 2.3 and 3.3 μm, a $CO_2$ feature at 2.7 μm, and $H_2O$ absorption at 2.7 and 3.7 μm. The reflected and emergent detectable fluxes in this region are very low for an Earth-like planet, making these features challenging to detect even for exoplanets orbiting the closest stars (**Figure 9**). In transiting geometry these features are slightly easier to detect (**Figure 10**).

In the IR, the 9.6 μm $O_3$ band is highly saturated and thus an excellent qualitative but poor quantitative indicator for the existence of $O_2$. The 7.66 μm $CH_4$ feature is hard to detect in low resolution for emergent flux for present-day Earth but is easily detectable at higher abundances, for example, on early Earth (**Figure 11**). The $CH_4$ is easier to detect in transmission (**Figure 10**). $N_2O$ features in the thermal IR at 7.75, 8.52, 10.65, and 16.89 μm can become observable for levels higher than in the present atmosphere of Earth. $CH_3Cl$ has spectral features between 6.5 and 7.5, 9.3 and 10.3, and 13 and 14.8 μm, which can potentially become detectable for active M stars (e.g., Segura et al. 2005b, Rauer et al. 2011, Rugheimer et al. 2015a) (**Figure 14**). Manufactured gases like CFCs ($CCl_2F_2$ and $CCl_3F$) also have absorption features in the IR but are undetectable at low resolution for Earth's atmosphere (Des Marais et al. 2002, Kaltenegger et al. 2007, Lin et al. 2014).

### 5.3. BIOSIGNATURES AND FALSE POSITIVES

The discussion about biosignatures in the literature is lively (for recent reviews, see, e.g., Kasting et al. 2014, Harman et al. 2015, Seager et al. 2016). Expanding sets of parameters are being explored for atmospheric models to assess which biosignatures could accumulate in a certain planetary environment and whether false positives could be produced by abiotic processes.

**5.3.1. Oxygen Alone as a Sign of Life?** Owen (1980) suggested searching for $O_2$ as a tracer of life. Present-day Earth's atmosphere consists of 21% oxygen. Oxygen is a reactive gas with a short atmospheric lifetime on Earth. Oxygenic photosynthesis, whose by-product is molecular oxygen extracted from water, allows terrestrial plants and photosynthetic bacteria to use abundant $H_2O$ instead of having to rely on scarce supplies of electron donors, like $H_2$ and $H_2S$, to reduce $CO_2$. Oxygenic photosynthesis by cyanobacteria and plants at a planetary scale results in the storage of large amounts of radiative energy in chemical energy, in the form of organic matter. Less than 1 ppm of atmospheric $O_2$ on present-day Earth comes from abiotic processes (Walker 1977). The reverse reaction, using $O_2$ to oxidize the organics produced by photosynthesis, can occur abiotically when organics are exposed to free oxygen or biotically by eukaryotes breathing $O_2$ and consuming organics. As discussed in Section 2.1.1, the net release of $O_2$ in Earth's atmosphere is due to the burial of organics in sediments (**Figure 3**) (see, e.g., the review by Kasting & Catling 2003). Each reduced carbon buried results in a free $O_2$ molecule in the atmosphere. This net release rate is balanced by the weathering of fossilized carbon when exposed to the surface. The oxidation of reduced volcanic gases such as $H_2$ and $H_2S$ also accounts for a significant fraction of the oxygen losses. The atmospheric oxygen is recycled through respiration and photosynthesis in less than 10,000 years. In the case of a total extinction of Earth's biosphere, the atmospheric $O_2$ would disappear in a few million years (Kaltenegger et al. 2009).

However, several teams have shown that oxygen can build up abiotically under certain geological settings (e.g., Schindler & Kasting 2000, Segura et al. 2007, Selsis et al. 2007, Leger et al. 2011, Hu et al. 2012, Wordsworth & Pierrehumbert 2013, Harman et al. 2015, Tian et al. 2014, Luger & Barnes 2015, Schwieterman et al. 2016). Some scenarios like oxygen buildup through photodissociation at the edges of or outside the HZ, for example, from water vapor photodissociation in a runaway greenhouse phase, would show other detectable features that could indicate oxygen's abiotic origin (see, e.g., Kasting 1988). However, any exoplanet scenario where lower or no $O_2$ sinks can be argued for would accumulate abiotically produced $O_2$, for example, after its creation from photodissociation by-products, with no clear way as yet to identify such cases as abiotic in low-resolution spectra. There has been lively discussion as to what kind of geological conditions would allow for such an organic buildup or whether these conditions would even be possible (see, e.g., Schaefer & Sasselov 2015). Therefore, not the existence of oxygen alone but the simultaneous presence of $O_2$ or $O_3$ and a reduced gas like $CH_4$ in a planet's atmosphere is strong evidence that the planet is inhabited (see Section 5.1).

**5.3.2. Methane and $N_2O$ as Signs of Life?** Approximately one-third of present-day $CH_4$ is produced via geological activity or methanogenic bacteria under anaerobic conditions on wetlands and in oceans (Kruger et al. 2001). The remaining two-thirds arises from human activity (industry and agriculture). A small fraction is produced abiotically in hydrothermal systems in which hydrogen is released by the oxidation of Fe by $H_2O$ and reacts with $CO_2$. Depending on the degree of oxidation of a planet's crust and upper mantle, such non-biological mechanisms can produce large amounts of $CH_4$ (e.g., Grenfell et al. 2010, Zendejas et al. 2010). Therefore, the detection of methane alone cannot be considered a sign of life, whereas its detection in an oxygen-rich atmosphere would be difficult to explain in the absence of a biosphere.

$N_2O$ on present-day Earth is produced in abundance by anaerobic denitrifying bacteria but only in negligible amounts by abiotic processes (e.g., Des Marais et al. 2002), and little is known about its past atmospheric levels. Energetic processes like lightning, UV radiation, and atmospheric shock from falling meteoroids can produce $N_2O$ (see, e.g., Kasting 1992, Zahnle et al. 2008). Models indicate buildup of abiotic $N_2O$ for an early Earth, assuming an active young Sun (Airapetian et al. 2016). Current levels of $N_2O$ would be hard to detect in Earth's atmosphere in low resolution (see, e.g., Segura et al. 2005, Selsis et al. 2006, Kaltenegger et al. 2007, Rauer et al. 2011). Its features are detectable in the IR, but they are located at wavelengths that correspond to the wing of the $CO_2$, $H_2O$, and $CH_4$ lines. Spectral features of $N_2O$ would become more apparent in atmospheres with more $N_2O$ and/or less $H_2O$ vapor (Segura et al. 2005b), making detection easier.

**5.4. EXTREME THERMODYNAMIC DISEQUILIBRIUM AS A SIGNATURE FOR LIFE?**
A potential biosignature of active, living processes is chemical disequilibrium in the surrounding environment. Lederberg (1965) suggested that the best remote signature of life would be extreme thermodynamic disequilibrium, like the combination of $O_2$ and $CH_4$. This concept has recently been explored (Bains & Seager 2012, Krissansen-Totton et al. 2016), but no additional, potentially observable disequilibrium redox pair has been identified. The appealing idea behind using it is that chemical disequilibrium would potentially be a generalized biosignature because it makes no assumptions about particular biogenic gases or metabolisms.

However, this idea cannot be generalized, as shown by a counterexample (e.g., Kasting 1992, Zahnle et al. 2008, Kasting et al. 2014): A planet's atmosphere with high concentrations of $H_2$ and CO would be out of thermodynamic equilibrium at room temperature because free energy considerations strongly favor the reaction $CO + 3\ H_2 \rightarrow CH_4 + H_2O$. Following Lederberg's (1965) idea, a simple interpretation would be that this disequilibrium was evidence of life.

However, CO-rich, $CH_4$-poor environments can be produced, for example, by photolysis of $CO_2$ in cold, dry, low- $O_2$ conditions or by impacts (Kasting et al. 2014). In addition, the presence of life would likely destabilize such a CO-rich atmosphere, as showcased by Earth's anaerobic biosphere, where models suggest that organisms would combine CO and $H_2$ to form $CH_4$, which at room temperatures would drive the atmosphere toward thermal equilibrium (e.g., Kharecha et al. 2005). CO is a high-free-energy compound that would be well out of thermodynamic equilibrium in almost any plausible abiotic terrestrial atmosphere (Harman et al. 2015). In addition, CO should be an excellent source of metabolic energy for microbes (Kharecha et al. 2005), and so its presence in a planet's atmosphere could be considered an antibiosignature (Zahnle et al. 2008). Although interesting, thermodynamic disequilibrium as a biosignature is hard to generalize; the specific life-detection criterion of the gas pair $O_2 + CH_4$ or $O_2 + N_2O$ proposed by Lovelock (1965) and Lippincott et al. (1967) is, however, very useful.

The thermodynamic chemical disequilibrium in the atmosphere of Solar System bodies, quantified by the available Gibbs free energy (Krissansen-Totton et al. 2016), shows that gas-phase disequilibrium in Earth's atmosphere is not unusual compared to other Solar System planet atmospheres and is smaller than that of Mars. Only when the authors take into account that Earth has an ocean can more significant changes between the Earth and other Solar System planets with atmospheres but without oceans be calculated, linking the usefulness of thermodynamic equilibrium to the knowledge of whether a planet has multiphase chemistry, for example, in the form of an ocean. As the authors point out, another very interesting open question is how these multiphase calculation values for Earth would compare to multiphase calculations for a second Solar System body, for example, Titan, and whether such calculations for an abiotic Earth would differ strongly from a biotic Earth.

### 5.5. SURFACE BIOSIGNATURES

Remote direct detection of surface life in reflected light becomes possible when organisms modify the detectable reflection of the surface (e.g., they influence the color of the surface). A ready example is vegetation on Earth. Vegetation has a specific reflection spectrum, with a sharp edge at approximately 700 nm, called the vegetation red edge (VRE). It is often suggested as a direct signature of life (e.g., Seager et al. 2005). Photosynthetic plants efficiently absorb the visible light but develop strong NIR reflection (possibly as a defense against overheating and chlorophyll degradation) resulting in a steep change in reflectivity. The primary molecules that absorb the energy and convert it to drive photosynthesis (of $H_2O$ and $CO_2$ into sugars and $O_2$) are chlorophyll A (0.450 μm) and B (0.680 μm). The exact wavelength and strength of the spectroscopic VRE depend on the plant species and the environment. Several groups (e.g., reviewed in Arnold 2008) have measured the integrated Earth spectrum via the earthshine technique, using sunlight reflected from Earth that is then reflected back by the Moon. Averaged over a spatially unresolved hemisphere of Earth, the additional reflectivity of this VRE feature is typically a few percent. Our knowledge of the reflectivity of different surface components on Earth, like deserts, oceans, and ice, helps in assigning the VRE of the earthshine spectrum to terrestrial vegetation. Minerals can exhibit a similar spectral shape at approximately 750 nm (e.g., Seager et al. 2005); therefore, the detection of the VRE on exoplanets, despite its interest, will not be unambiguous. Assuming that similar photosynthesis would evolve on a planet orbiting other host stars, the VRE could be shifted to different wavelengths (Kiang et al. 2007a,b). A thorough analysis of the likelihood of oxygenic photosynthesis arising elsewhere is given by, for example, Rothschild (2008).

Vegetation is only one among many surface features life produces on Earth (see, e.g., Kiang et al. 2007a, Cockell et al. 2009, Hu et al. 2012, Sanroma et al. 2013, Hegde et al. 2015). Land plants have been widespread on Earth for only approximately 460 Myr (e.g., Zahnle et al. 2007), whereas much of the history of life has been dominated by single-celled microbial life. Biota on Earth generates a wide range of characteristic reflectivity and colors. To mimic detectable surface reflection features of known life, Hegde et al. (2015) measured the spectral characteristics of 137 phylogenetically diverse microorganisms containing a range of pigments, including ones isolated from Earth's most extreme environments. The team used an integrating sphere, which mimics the observing geometry of an exoplanet that is modeled as a Lambertian sphere. This database provides high resolution hemispherical reflectance measurements for the visible and NIR (0.35–2.5 μm) spectra for a subset of life known on Earth and is freely available *(http://carlsaganinstitute.org/data)*. In this subset are extremophiles that provide us with the minimum known envelope of environmental limits for life on our planet (Rothschild & Mancinelli 2001). A library capturing the range of possible reflectivity of biota on Earth as it would be seen on exoplanets is critical to inform disk-integrated observations and models of rocky exoplanets.

Planets that receive high doses of UV radiation are generally considered to be less promising candidates in the search for life. During M-star flares, the UV flux on a planet in the HZ can increase by up to two orders of magnitude, dramatically increasing the planet's surface radiation environment (e.g., Segura et al. 2010). In spite of the periodically high UV fluxes, several teams have made the case that planets in the HZ of M stars can remain habitable (e.g., Heath et al. 1999, Buccino et al. 2006, Scalo et al. 2007, Segura et al. 2010, O'Malley-James & Kaltenegger 2017a,b).

On Earth, biological mechanisms such as protective pigments and DNA repair pathways can prevent or mitigate radiation damage. Additionally, subsurface environments can reduce the intensity of radiation reaching an organism (e.g., Ranjan & Sasselov 2016). Life that is constrained to habitats underwater or beneath a planet's surface may not be detectable remotely. However, another possible response of the biosphere is detectable, photoprotective biofluorescence, in which protective proteins absorb harmful UV wavelengths and re-emit them at longer, safer wavelengths. This could be an alternative UV protection mechanism that generates additional visible flux coupled to the incoming stellar UV. On Earth, evidence suggests that some coral species use photoprotective biofluorescence as a mechanism to reduce the risk of damage to symbiotic algae, which provide the coral with energy (e.g., O'Malley-James & Kaltenegger 2017a): Fluorescent proteins in corals absorb blue and UV photons and re-emit them at longer wavelengths. For planets orbiting active stars, using such a defense mechanism would not only allow life to still use the surface environment but also generate a new detectable biosignature. Such a signal could be comparable in strength to reflection of vegetation on Earth, but unlike vegetation reflection that is a specific fraction of the incoming visible light, biofluorescence would temporarily generate additional visible planetary flux. It could be a strong temporal biosignature on planets orbiting, for example, active M stars, like Proxima Centauri. The resulting UV flare–induced biofluorescence could even uncover normally hidden biospheres during a flare.

### 5.6. DAILY LIGHT CURVES

Daily changes in surface reflectivity or IR flux could identify different surface features, like continents and oceans, for cloud-free planets (e.g., Ford et al. 2001, Gaidos & Williams 2004). On a cloud-free Earth, the diurnal flux variation in the visible spectrum caused by different surface features rotating in and out of view could be high, but only if a planet exhibits hemispheric inhomogeneity. In principle, principal component analysis of time variations in the multiband light curves for Earth could distinguish terrains with significantly different reflectivity, including oceans, clouds, and land surface (e.g., Fujii et al. 2011, Gomez-Leal et al. 2012, Cowan & Strait 2013). Different dynamics regimes on exoplanets will influence whether such techniques are possible, but with large telescopes and high temporal multiband observations, exoplanet characterization will become even more intriguing.

When a planet is only partially illuminated, a more concentrated signal from surface features could be detected as they rotate in and out of view on a cloudless planet. However, the overall detectable signal in reflected light decreases significantly if only a small part of the planet's lit surface is seen, making this an effect that can only be addressed with future telescopes that can collect enough light when only seeing a small fraction of the planet's surface. One such proposed signature is an ocean glint (e.g., Williams & Gaidos 2008, Robinson et al. 2014). False positives for such an ocean glint could be generated depending on a planet's obliquity (Cowan et al. 2012).

### 5.7. How Clouds Change the Picture

Surface features can only be detected if the planet's atmosphere is transparent and the surface reflects the incoming light. For planets with greater land and vegetation coverage as well as less cloud coverage than Earth, detection of surface features would become easier. With greater cloud coverage and less surface coverage, detection would become harder. The depth of spectral features in reflected light is dependent on the abundance of a chemical as well as the incoming stellar radiation at that wavelength. In thermal emission, the depth of spectral features depends on the abundance of a chemical as well as the temperature difference between the emitting/absorbing layer and the continuum radiation of the planet.

The height and other properties of the cloud (or hazes) will influence whether the atmospheric signatures will be easier or harder to detect (e.g., Kaltenegger et al. 2007, Rauer et al. 2011, Zsom et al. 2012, Rugheimer et al. 2013, 2015b, Arney et al. 2016). High-reflectivity clouds increase the reflectivity of an Earth-like planet in the visible to near IR substantially and therefore can increase the equivalent width of all observable features even though they block access to some of the lower atmosphere. In the IR, these clouds influence the continuum emission of the planet, making some features easier to detect and some harder, depending at what temperature they absorb or emit compared to the continuum temperature. Clouds influence a planet's climate and can also shift the limits of the HZ as discussed above (see e.g. Kopparapu et al. 2014).

Earth, Venus, and to a small extent Mars all have clouds; therefore, we expect that exoplanets in the HZ also harbor clouds in their atmospheres. Highly reflective surfaces like clouds, or ice and snow, increase the brightness of an Earth-like planet in the visible to NIR and decrease its thermal flux due to a lower emitting temperature, making frozen or cloudy planets easier to detect at visible wavelengths.

Exploring surface features of Earth-like planets becomes possible if either no significant cloud cover exists on an exoplanet or the SNR of each observation is sufficiently high to remove the cloud contribution from the overall detected signal (e.g., Palle et al. 2008, Cowan et al. 2009, Fujii et al. 2011) assuming that the clouds move and their movement is not linked to surface rotation. Several observations per rotation period would be needed to distinguish clouds from surface features that are bound to the rotation of the planet (Palle et al. 2008); the same study showed that using this approach, one could retrieve the planetary rotation rate. These results are intriguing. However, giving enough SNR to peer at the surface of partially cloud-covered habitable planets will require telescopes more powerful than the *James Webb Space Telescope* ( JWST) and the ELTs.

### 5.8. Evolution of Biomarkers Over Geological Time on Earth

Observations of terrestrial exoplanet atmospheres will occur for planets at different stages of geological evolution. Earth's atmosphere has experienced dramatic evolution over 4.5 Gyr (e.g., Ribas et al. 2005, Zahnle et al. 2008). As discussed in Section 2.3, a star brightens with time; therefore, early Earth only received approximately 70% of the current solar irradiation when it was formed, which increased gradually over time as the Sun brightened. Such low solar irradiant flux should have led to subfreezing global temperatures for the first 1–2 Gyr assuming an atmospheric composition similar to that for the present day. In spite of this, temperatures must have been higher under the less luminous Sun, or geological records would show a frozen young

Earth. This apparent conundrum is what is often deemed the faint young Sun paradox (e.g., Sagan & Mullen 1972, Goldblatt & Zahnle 2011, Kasting 2010).

Geological records show that Earth's atmosphere changed significantly over geological time, balancing the lower initial solar irradiance (reviewed in Kasting & Catling 2003). Therefore, Earth's spectrum also changed throughout its geological evolution due to variations in chemical makeup, temperature structure, and surface morphology over time.

Earth's spectra at different geological stages (e.g., Pavlov et al. 2000, Schindler & Kasting 2000, Selsis 2000, Traub & Jucks 2002, Segura et al. 2003, Meadows et al. 2005, Kaltenegger et al. 2007, Domagal-Goldman et al. 2011, 2014, Arney et al. 2016) provide additional data points for comparison with potentially habitable exoplanets. At approximately 2.3 Ga, oxygen and ozone became abundant, affecting the atmospheric absorption component of the spectrum significantly. The rise of oxygen made the biosignature pair of oxygen or ozone in combination with a reduced gas, methane, detectable in the IR and NIR (Kaltenegger et al. 2007).

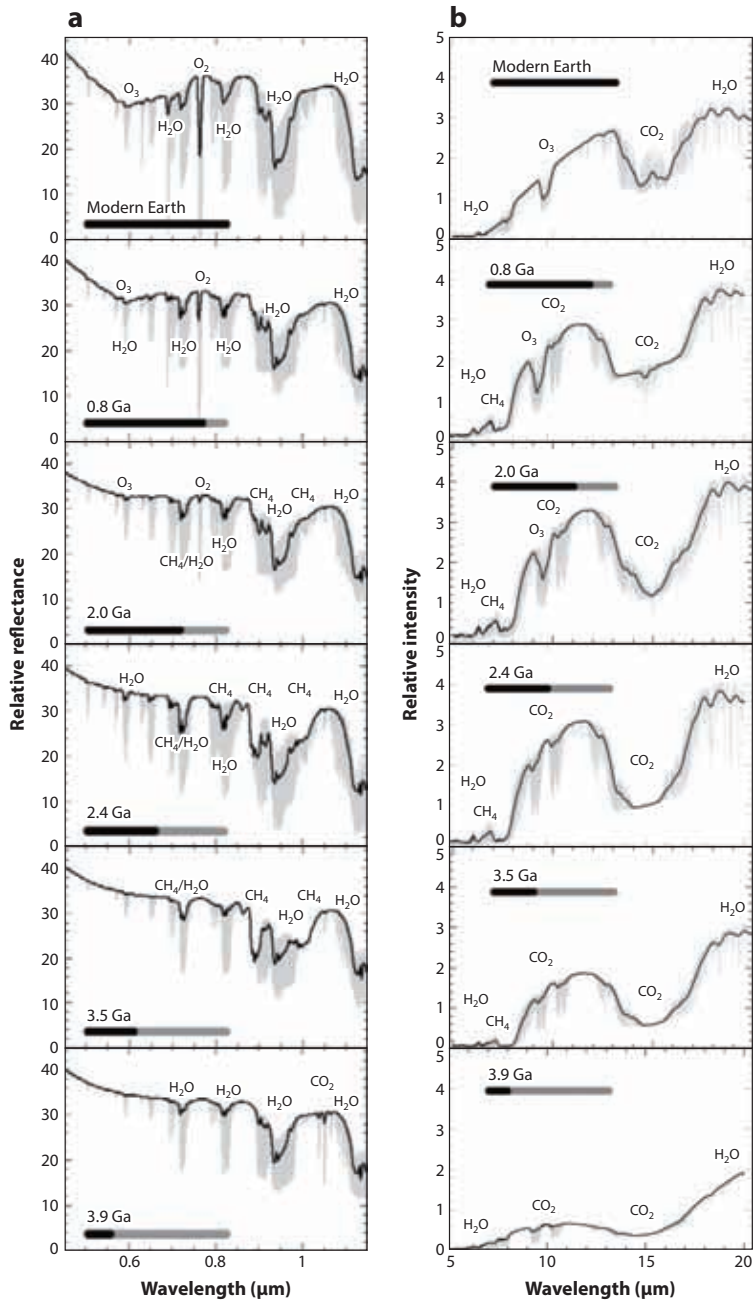

**Figure 11** Changes in Earth's atmospheric composition through geological times influence its spectrum: (*a*) visible to NIR, and (*b*) IR spectral features on an Earth-like planet show considerable changes. Emergent visible to IR spectra of Earth at six major developmental epochs during its geological evolution from a $CO_2$-rich (3.9 Ga) to a $CO_2/CH_4$-rich atmosphere (2 Ga) to a present-day atmosphere (modern Earth). Bold lines show spectral resolution ($\lambda/\Delta\lambda$) of 80 (visible to NIR) and 25 (IR), and high-resolution spectra are depicted in light gray. Abbreviations: IR, infrared; NIR, near-infrared. Data from Kaltenegger et al. (2007).

The spectrum of the Earth has exhibited a strong IR signature of ozone and a weaker signature for methane in the IR for more than 2 Gyr (**Figure 11**), and a strong visible signature of $O_2$ combined with a weak NIR signal for methane for an undetermined period of time between 2 and 0.8 Ga (**Figure 11**), depending on the required depth of the band for detection and also the actual evolution of the $O_2$ level. This difference is due to the fact that a saturated ozone band already appears at very low $O_2$ levels ($10^{-4}$ ppm), whereas the oxygen line remains unsaturated at values below current levels (Leger et al. 1993, Segura et al. 2003). In addition, the stratospheric warming decreases with the abundance of ozone, making the IR $O_3$ feature deeper because of the increased temperature difference between the surface-cloud continuum and the ozone layer. The methane features are more pronounced at earlier epochs in both the IR and NIR spectra. $CH_4$ shows weaker detectable signatures in the NIR and IR throughout geological evolution. For the past 450 Myr, an extensive land plant cover generated a shallow red chlorophyll edge in the reflection spectrum of our planet (as discussed in Section 5.7).

**Figure 11** shows emergent visible and IR spectra models of Earth at six major developmental epochs during its geological evolution (Kaltenegger et al. 2007). The atmospheres shown in **Figure 11** correspond to a prebiotic Earth model at 3.9 Ga and 3.4 Ga, a model atmosphere for the early rise of oxygen at 2.4 Ga and 2.0 Ga, moderate to current levels of oxygen at 0.8 Ga, and modern Earth, respectively, and show that different atmospheric signatures are observable through Earth's evolution.

As shown in **Figure 11**, the presence on Earth of biogenic gases such as $O_2/O_3$ in combination with $CH_4$ should have been detectable remotely since approximately 2 Ga in the IR (with a resolution of 25) and between 0.8 and 2 Ga in the visible (with a resolution of 100) spectra, depending on the oxygen buildup (Kaltenegger et al. 2007). These features imply the presence of an active biosphere, although their absence does not necessarily imply the absence of life, as shown by its evolution on our planet. Life existed on Earth before the interplay between oxygenic photosynthesis and carbon cycling produced an oxygen-rich atmosphere, but it did not leave a unique, remotely observable signature.

If an extrasolar planet were found with a corresponding spectrum, we could use the evolutionary stages of our planet to characterize it in terms of habitability and the degree to which it showed signs of life. Discussions about the exact levels of oxygen through geological time are ongoing (see, e.g., Holland 2006, Lyons et al 2014, Planavsky et al. 2014). We can learn about the evolution of our own planet's atmosphere and possibly the environment needed for the emergence of life by observing exoplanets in different evolutionary stages.

**5.9. Earth's Surface Ultraviolet Environment Through Geological Time for Different Host Stars**

Depending on the intensity, UV radiation can be both useful and harmful to life as we know it: UV radiation from 180 to 300 nm can inhibit photosynthesis and cause damage to DNA and other macromolecules (Voet et al. 1963, Matsunaga et al. 1991, Tevini 1993, Kerwin & Remmele 2007). explore how much UV flux reached the surface of Earth. However, these same wavelengths also drive several reactions thought necessary for the origin of life (e.g., Senanayake & Idriss 2006, Ritson & Sutherland 2012, Ranjan & Sasselov 2016). Models of the UV surface environments for atmospheres that correspond to geological epochs throughout Earth's evolution (Segura et al. 2005b, Rugheimer et al. 2015b, Arney et al. 2016).

**Figure 12** shows model surface UV flux for three epochs of Earth's history: a prebiotic Earth model at 3.9 Ga, a model atmosphere for the early rise of oxygen at 2.0 Ga, and modern Earth (see Rugheimer et al. 2015b for details).

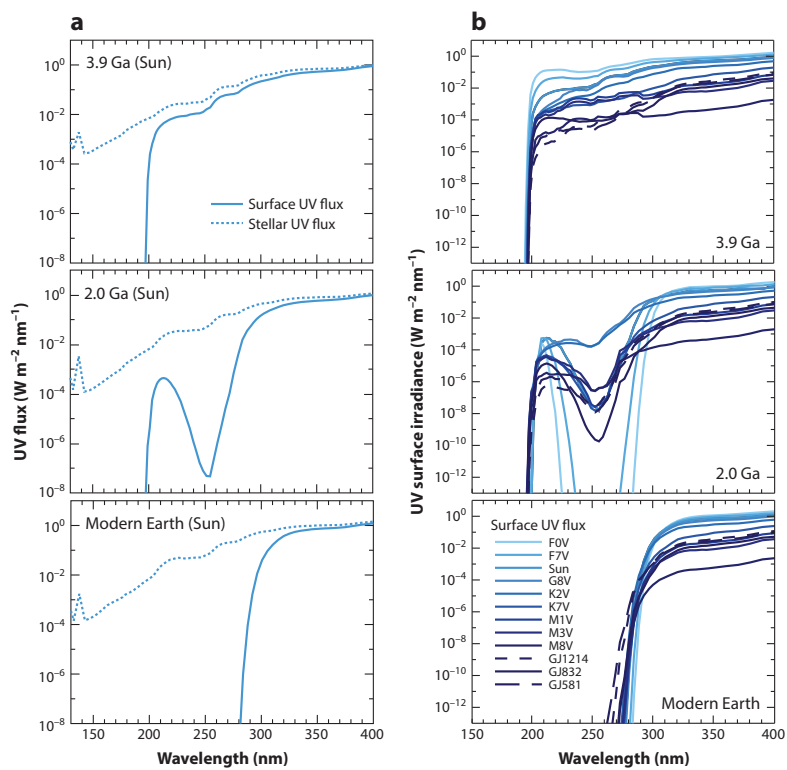

**Figure 12** Stellar and model surface UV flux for (*a*) Sun–Earth for three geological epochs show how the UV environment changed throughout Earth's history from a prebiotic Earth (3.9 Ga) to a low-oxygen environment (2.0 Ga) to modern Earth. (*b*) Using outgassing rates for the Earth model through geological times for planets that orbit different host stars shows how the surface UV environment changes due to the host star. Data from Rugheimer et al. (2015b).

For prebiotic atmospheric models, a significant portion of the high-energy, incident UV-C flux reached Earth's surface. With higher ozone concentration in the atmosphere, that value decreases. Hazes or a layer of water or soil can shield life from UV radiation (Cockell 1998, Cnossen et al. 2007, Arney et al. 2016, Ranjan & Sasselov 2016). Recent studies have explored how much UV flux life would have to cope with on the surface of Earth-like planets orbiting other host stars for atmospheres corresponding to Earth's geological evolution (Rugheimer et al. 2015b, O'Malley-James & Kaltenegger 2017b).

### 5.10. Biosignatures for Planets Orbiting Different Host Stars

The UV environment of a host star dominates the photochemistry, which influences the resulting atmospheric constituents and biosignatures for terrestrial planets (e.g., Selsis et al. 2002, Segura et al. 2003, 2005, Grenfell et al. 2007, Rauer et al. 2011, Hedelt et al. 2013, Rugheimer et al. 2013, 2015b, Tian et al. 2014, Domagal-Goldman et al. 2014). The strength of the atmospheric absorption features varies significantly between planets orbiting different host stars with spectral types for both transmission (**Figure 13**) and emergent (**Figure 14**) spectra because the SED of the host star influences the atmospheric chemical makeup of an Earth-like planet. Features like $CH_4$ and $CH_3Cl$ should be easier to detect for Earth-analog planets around cool host stars (see, e.g., Segura et al. 2003, Rauer et al. 2011, Rugheimer et al. 2013, 2015a,b).

To date, few observations exist in the UV region for M dwarfs. The Hubble UV program to observe active M dwarfs, MUSCLES (Measurements of the Ultraviolet Spectral Characteristics of Low-mass Exoplanetary Systems; France et al. 2013, Youngblood et al. 2016), has expanded our knowledge of the effect of M-star spectra on atmospheric models and biosignature detection significantly even though it only targeted six stars. Further UV spectra for cool stars are critical to explore their influence on detectable biosignatures.

In the visible to NIR spectra, the strongest atmospheric features from 0.4 to 4 μm for Earth-like planets orbiting F- to M-type stars are $O_3$ at 0.6 μm, $O_2$ at 0.76 μm, $H_2O$ at 1.9 μm, and $CH_4$ at 1.7 μm, as shown in **Figure 13**.

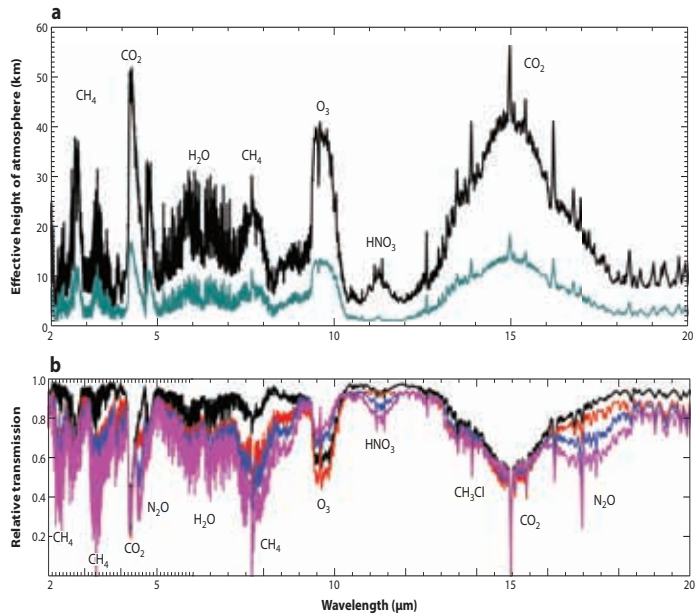

**Figure 13** (*a*) Effective heights: Earth (*black*) compared with a super-Earth with three times Earth's gravity (*green*) around the Sun. (*b*) Comparison of atmospheric transmission of an Earth-analog planet around the Sun (*black*), the active M3.5V star AD Leo (*red*), and two inactive M-star models, M5V (*blue*), and M7V (*magenta*). Data from H. Rauer, DLR.

The depth of the $H_2O$ absorption features increases for planets orbiting cool stars due to their increased $H_2O$ abundance. The depth of the $CH_4$ absorption feature increases with decreasing stellar $T_{eff}$ due to the increase in $CH_4$ abundance. Whether $O_2$ can be observed at lower concentrations in emergent spectra will depend largely on clouds, with possible degeneracies between $O_2$ concentration and percent cloud cover. The relative depth of the $O_2$ feature at 0.76 μm is increasingly difficult to detect in reflected light of later M dwarfs owing to low stellar fluxes in that wavelength region.

In the IR, the strongest atmospheric features from 4 to 20 μm for Earth-like planets orbiting F- to M-type stars are $O_3$ at 9.6 μm, $CO_2$ at 15 μm, $H_2O$ at 6.4 μm, and $CH_4$ at 7.7 μm, as shown in **Figures 9** and **13**. In emergent and transmission spectra, the $CH_4$ feature is prominent in the planetary spectra around cool stars owing to high $CH_4$ abundance in low-UV environments. $N_2O$ builds up to observable concentrations in planetary models around M dwarfs with low UV flux. $CH_3Cl$ could also become detectable, depending on the depth of the overlapping features. In emergent flux spectra, the depth of the $O_3$ feature decreases for hotter stars, despite increasing $O_3$ abundance, owing to lower contrast between the continuum and absorption layer temperatures. $O_3$ appears in emission for $T_{eff} \geq 6,500$ K because of the lower continuum temperature

The observability of the biosignature gases, $O_2/O_3$ in combination with $CH_4$, reduces with increasing cloud cover and increases with planetary age for MS host stars from F0 to M8 (Rugheimer et al. 2013). **Figure 13** also shows the model transmission spectra of a present-day Earth-analog planet compared with a super-Earth with three times Earth's gravity (Rauer et al. 2011), in which the increased gravity of a super-Earth in these models compresses the effective height of the planet's atmosphere, making spectral features for an Earth-analog super-Earth potentially challenging to detect. However, if the outgassing rates per surface area in the model are kept constant while the surface pressure increases with gravity, the effective height of super-Earth atmospheric features would also be similar to that of Earth-analog planets (Kaltenegger et al. 2009). If the surface pressure does not scale with gravity but is lower, then the effective height due to atmospheric features on super-Earths would be larger and easier to detect.

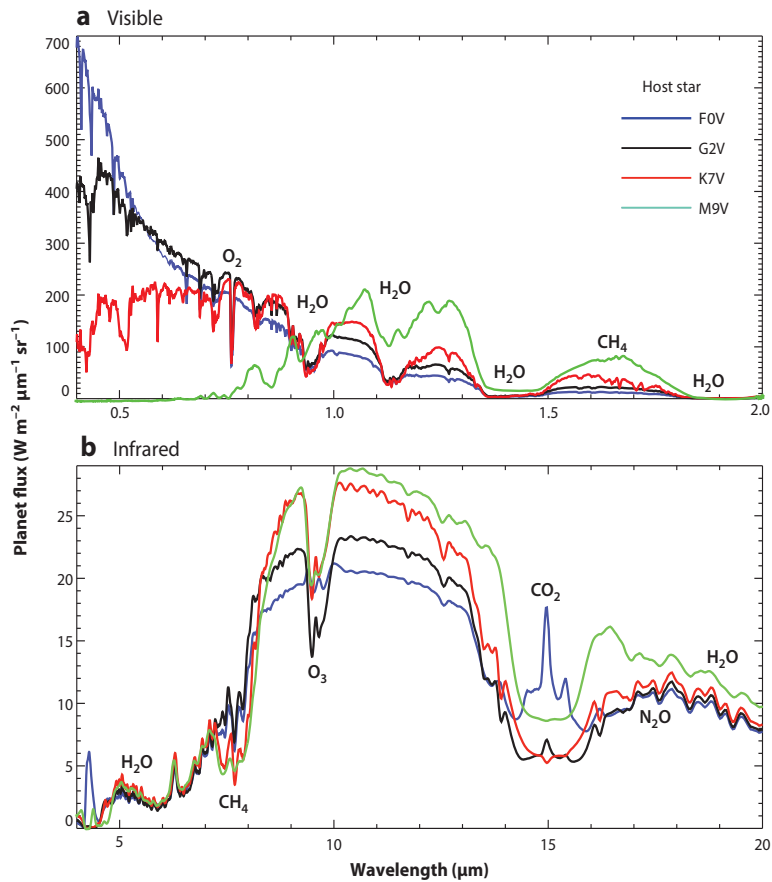

**Figure 14** Atmospheric features in the (*a*) visible and (*b*) infrared for Earth-like planets orbiting different host stars. Data from Rugheimer et al. (2013, 2015a).

## 6. OUTLOOK

Our closest star, Proxima Centauri, a cool M5V dwarf only 1.3 pc from the Sun, harbors a planet in its HZ (Anglada-Escude et al. 2016). In the close-by TRAPPIST-1 planetary system, three to four Earth-size planets orbit in the HZ of its M9V host star only about 12 pc from the Sun (Gillon et al. 2017). These two planetary systems already show several interesting targets for the future characterization of potentially habitable worlds orbiting neighboring stars.

### 6.1. Cool Red Stars as Interesting Targets

Cool red stars, M stars, are the most common type of star in the galaxy and make up 75% of the stars in the solar neighborhood. They are also excellent candidates for HZ terrestrial planet searches due to the increased transit probability, transit frequency, and higher SNR for both transit and radial velocity detections at a given planetary size and mass. But they are not just good targets for observations; first estimates of the occurrence rate of Earth-sized planets in the HZ around cool dwarfs are very high, ranging between 15% and 66% (**Table 1**), boosting speculations that the Universe could be teeming with life on planets orbiting red stars.

Several potential drawbacks for planets in the HZs of M stars have been discussed, creating a lively field of research (e.g., reviewed in Scalo et al. 2007, Shields et al. 2016). One concern is that the habitability of planets orbiting M-dwarf stars is complicated by the small orbital distance of the HZs of these stars, due to their low stellar luminosities. The strong tidal interaction between a planet in the HZ of an M star and its host star will lead to a potentially tidally locked or synchronously rotating planet. If other planets exist in the system, synchronous locking is unlikely, but the rotation rate of the planet could be reduced due to the tidal locking. This will in turn influence the atmospheric dynamics and cloud coverage, and the planet's ability to generate a strong magnetic field, which could strip its atmosphere. However, even if a planet were synchronously rotating, several studies have shown that if the planet has an atmosphere similar to that of Earth (1 bar surface pressure) and a body of water, an ocean and/or a substantial atmosphere can efficiently transport heat from the day-side to the night (e.g., Joshi 2003, Yang et al. 2014). Very slow rotating planets with higher surface pressure could therefore be better environments for habitability than planets with lower surface pressures.

As discussed in Section 2.2, the light of a cooler star is more efficient in heating a planet. Surfaces such as ice also reflect less light at longer wavelengths, at which most of the light from a cool M star will be emitted, reducing the planetary albedo further. This lowered ice albedo can also keep a planet from going into a snowball state, due to the reduced effectiveness of the ice albedo feedback on such a planet (e.g., Joshi & Haberle 2012, Shields et al. 2013, Von Paris et al. 2013), potentially increasing the stability of its climate. Note that this effect will not extend the limits of the HZ outward, because at the edges of the HZ, the albedo of a planet is dominated by the atmosphere not by the surface.

One serious concern is the activity of the host star. X-ray and EUV flare activity can occur up to 10–15 times per day, and typically 2–10 times, for young M dwarfs (France et al. 2013, Cuntz & Guinan 2016), which increases atmospheric erosion on close-in planets. This should in turn result in higher fluxes of UV radiation reaching the planet's surface and, potentially, a less dense atmosphere (e.g., Lammer et al. 2007). In addition, planets in the HZs of M stars are subject to stellar particle fluxes that are orders of magnitude stronger than those in the solar HZ (Cohen et al. 2014) and could erode the planet's protective ozone shield as well as some of the atmosphere. M stars also remain active for longer periods of time than the Sun (e.g., West et al. 2011), making the surface of planets around an active M star a potentially highly irradiated environment, depending on the atmospheric composition. Life could shelter subsurface on such planets, for example, in an ocean, which would make it harder to detect remotely. If photosynthetic life developed on a highly UV-irradiated planet, it could employ UV defenses, such as living under soil layers or underwater. However, there may be other detectable surface biosignatures that would be associated with high-UV surface environments. Some biological UV-protection methods, such as biofluorescence, could make such a biosphere detectable (O'Malley-James & Kaltenegger 2017a).

The question of a planet's habitability is also linked to its ability to acquire and retain water. Arguments against M-star planet habitability focused on the difficulty that planetesimals in M-star HZs would have in acquiring volatiles. This was attributed either to higher orbital speeds, leading to more energetic collisions with other bodies (Lissauer 2007), or to starting out dry because of inefficient radial mixing, and so fewer volatile-rich planetesimals from greater distances would be accreted (Raymond et al. 2007). These concerns have been questioned by, for example, the in situ accretion model of Hansen (2014), which predicts that gatekeeper planetesimals may be large enough to reduce collisional velocities, allowing M-star planets to acquire many hundreds of times Earth's surface water endowment. As mentioned in Section 2.2, accreting planets that are later located in the MS HZ orbiting host stars of stellar type K5 and cooler receive stellar fluxes that exceed the runaway greenhouse threshold, and thus may lose a substantial part of the water initially delivered to them (Ramirez & Kaltenegger 2014, Luger & Barnes 2015). Therefore, M-star planets in the HZ need to initially accrete more water than Earth did or, alternatively, have additional water delivered later to remain habitable.

### 6.2. A Close-By Test Case – Proxima b

Our closest neighboring star, Proxima Centauri, is an active M5 star that currently experiences intense flares every 10–30 hours (Cincunegui et al. 2007) and shows a radial velocity signal that indicates a terrestrial 1.3-$M_\oplus$ planet in its HZ at an 11.7-day orbit (Anglada-Escude et al. 2016). It receives a stellar flux equivalent to approximately 0.65 times the solar flux on Earth. However,

Proxima Centauri is a flare star that flares strongly. Proxima b receives 30 times more EUV than Earth and 250 times more X-ray radiation (Ribas et al. 2016). Without further observational constraints, a variety of atmospheres are possible for Proxima b (e.g., Meadows et al. 2016, Ribas et al. 2016, Turbet et al. 2016).

Given the proximity of Proxima b to its host star, these EUV and X-ray fluxes place Proxima b's atmosphere at risk of significant erosion. In the absence of a strong planetary magnetic field, the result would be a thinner atmosphere and/or lower atmospheric ozone levels, which would allow more of Proxima Centauri's UV flux to reach the planet's surface. These conditions make Proxima b a very interesting planet, because if we were to find life on such a planet, it would prove the universal tenacity of life.

### 6.3. FUTURE MISSIONS: A SHORT PREVIEW

The favorable contrast ratio between a cool host star and its potentially habitable rocky planet makes such planets interesting targets for upcoming ground- and space-based observations. Near future all-sky survey missions like the *Transiting Exoplanet Survey Satellite* (TESS) to be launched in 2018 (Ricker et al. 2014) and precision photometry survey instruments like the PLAnetary Transits and Oscillations of stars (PLATO) to be launched in 2024 (Rauer et al. 2014) as well as the European *CHaracterising ExOPlanets Satellite* (CHEOPS) to be launched in 2018 (Fortier et al. 2014) are expected to provide further data and insight into the characteristics and occurrence rate of small planets around nearby and bright stars. CHEOPS will look for transits of already-detected RV exoplanets for which the mass is known.

TESS will survey the whole sky to identify transiting exoplanets around nearby and bright stars, including terrestrial exoplanets in the HZ of cool stars. It will be sensitive enough to identify HZ planet candidates around a large number of nearby low-mass stars ($T_{eff} \leq 4,000$ K; late M and early K stars) for future ground- and space-based characterization. TESS is expected to find hundreds of 1.25–2-$R_\oplus$ planets and tens of Earth-sized planets, with a handful (<20) of these planets in the HZ of their host stars (Sullivan et al. 2015). CHEOPS could also observe selected TESS targets in addition to known RV planets to provide a longer observation baseline for selected targets. For transiting terrestrial planets around the closest stars, the JWST scheduled for launch in late 2018 (e.g., Gardner et al. 2006, Clampin et al. 2009, Deming et al. 2009, Kaltenegger & Traub 2009, Barstow & Irwin 2016) as well as upcoming ground-based telescopes (e.g., Broeg et al. 2013, Snellen et al. 2013, Rodler & Lopez-Morales 2014) might be able to detect biosignatures in a rocky planet's atmosphere.

Space mission concepts to characterize Earth-like planets are currently being designed, for example, by NASA's science and technology definition teams, but no concept has been selected yet. Different concepts like star shades, coronagraphs, or large stable UV to VIS and IR telescopes are designed to take spectra of extrasolar planets with the ultimate goal of remotely detecting atmospheric signatures to characterize nearby super-Earths and Earth-like planets, enable comparative planetology beyond our Solar System, and search for signs of life on other worlds.

### 6.4. DATABASE OF EXOPLANET SPECTRA

The first detections of EGP spectra are already giving us a glimpse into the fascinating diversity of extrasolar planets. Rocky exoplanets show a wide range of properties (**Figure 1**). We are learning how to retrieve parameters like molecular abundance from transiting as well as directly imaged EGPs and mini-Neptunes, with the intention to use these methods on rocky planets when their spectra become available with the next generation of space- and ground-based telescopes.

We use models based on Earth as a Rosetta stone to explore how we could detect habitats outside our Solar System orbiting their host stars or planets. Other rocky bodies in our Solar System, especially Mars and Venus, help establish some boundaries on habitable conditions and inform atmospheric models for very different stellar irradiance than Earth's. One-dimensional models that model a planet's disk-integrated properties can explore a wide parameter range of environments. Atmospheres of extrasolar planets will allow us to explore different planetary environments remotely. A spectral database for a wide range of habitable world models (see, e.g., the freely available spectral database at *http://carlsaganinstitute.org/data*) can inform initial observation strategies and interpretation of observations. Such models will be refined once observed spectra are available, helping us understand what environments we are observing.

For very different planetary conditions, it is important to combine the broad-parameter space exploration of 1D models with modified 3D models that eliminate empirical parameters derived for Earth's atmosphere in favor of the underlying physics that can be scaled to other exoplanets (as shown for the HZ boundaries). Such 3D models are needed to inform where simplified 1D models can be used and where 3D models are needed, as for very slow rotating planets that could develop a very different cloud structure or thin atmospheres that might not transfer heat efficiently to the night side of a planet. Even though most parameters of potentially habitable extrasolar planets are still unknown, modeling a large grid of planetary environments and their detectable spectra already allows us to explore which parameters will influence remotely detectable features. Only the planetary parameters, which change the observable spectrum, can be constrained with remote observations.

Finding thousands of exoplanets has taken the field of comparative planetology beyond the Solar System. We know that planets orbiting other stars are common. The Kepler data show that approximately every fifth star hosts at least one rocky planet in its HZ. We have also already identified the first few dozen host stars of such planets. Determining how to identify habitable environments remotely is the next challenge. Such remotely detectable spectra will give us insights into how bodies that may be able to sustain habitable conditions, like Earth, evolve and what conditions allow for the origins of life.

## 7. SUMMARY

The search for rocky planets has revealed a fascinating diversity of worlds, and with the upcoming generation of telescopes like the JWST and the E-ELTs, we should be able to explore such worlds remotely. We can probe for signs of life, pushing the limit of technical capabilities. Any information we collect on habitability is embedded in a context that allows us to interpret what we find and to test and refine 1D and 3D atmospheric models. To search for signs of life, we need to understand how an observed atmosphere physically and chemically works. Observations of rocky exoplanets will give us insights into these fundamental questions, improving our understanding of how habitable worlds function.

The emerging field of extrasolar planetary search has shown an extraordinary ability to combine research from astrophysics, chemistry, biology, and geophysics in a new and exciting interdisciplinary approach to understand our place in the Universe. And at the limit of technical capability, using Earth as a Rosetta stone for habitable planets, it might soon revolutionize our worldview again with the detection of signs for habitability on other worlds. In a few years, we may have the first spectra for rocky exoplanets in the HZ of their host stars and will be able to start comparative planetology beyond the Solar System with dozens to hundreds of potential habitable worlds.

**DISCLOSURE STATEMENT** The author is not aware of any affiliations, memberships, funding, or financial holdings that might be perceived as affecting the objectivity of this review.

**ACKNOWLEDGMENTS** The volume of literature on this field is exciting, showing the progress we are making one these fundamental questions to figure out how to characterize other Earth-like planets. It also engages in a lively discussion on different topics, which moves our understanding forward and invigorates our search. However given the page limit of this review, I could not cover all the topics and aspects I wanted to cover. I therefore sincerely apologize to all those colleagues whose work may not have received due mention in this review. I am deeply grateful to Ramses Ramirez, Jack Madden, Siddharth Hegde, Jack O'Malley-James, and Yamila Miguel from the Carl Sagan Institute (CSI) at Cornell University for in depth discussions and insightful constructive comments. I want to acknowledge the enthusiastic help I received in generating the figures by Li Zeng from Harvard University for generating **Figure 1**, Franck Selsis from Bordeaux Observatory for generating **Figure 3**, Ramses Ramirez from CSI for generating **Figures 4** and **5**, and Jack O'Malley-James from CSI for compiling **Table 2**. The author acknowledges support by the Simons Foundation (SCOL 290357, L.K.) and the Carl Sagan Institute at Cornell University.

## LITERATURE CITED

Abbot DS. 2016. *Ap. J.* 827:117

Abbot DS, Cowan NB, Ciesla FJ. 2012. *Ap. J.* 756(2):178

Abe Y, Abe-Ouchi A, Sleep NH, Zahnle KJ. 2011. *Astrobiology* 11:5

Agol E, Jansen T, Lacy B, et al. 2015. *Ap. J.* 812:5

Agol E, Steffen J, Sari R, Clarkson W. 2005. *MNRAS* 359:567

Airapetian VS, Glocer A, Gronoff G, et al. 2016. *Nat. Geosci.* 9:452–55

Anglada-Escudé G, Amado PJ, Barnes J, et al. 2016. *Nature* 536:437–40

Arney G, Domagal-Goldman SD, Meadows VS, et al. 2016. *Astrobiology* 16(11):873–99

Arnold L. 2008. *Space Sci. Rev.* 135:323–33

Bains W. 2004. *Astrobiology* 4:137–67

Bains W, Seager S. 2012. *Astrobiology* 12:271–81

Barnes JW, O'Brien DP. 2002. *Ap. J.* 575:1087

Barnes R, Jackson B, Greenberg R, Raymond SN. 2009. *Ap. J. Lett.* 700(1):L30–33

Baross J, Comm. Limits Org. Life Planet Syst., Comm Orig. Evol. Life, et al. 2007. *The Limits of Organic Life in Planetary Systems*. Washington, DC: Natl. Acad.

Barstow JK, Irwin PGJ. 2016. *MNRS Lett.* 461(1):L92–96

Batalha NM. 2014. *PNAS* 111(35):12647–54

Berner RA, Lasaga AC, Garrels RM. 1983. *Am. J. Sci.* 283:641–83

Betremieux Y, Kaltenegger L. 2013. *Ap. J. Lett.* 772:L31

Betremieux Y, Kaltenegger L. 2014. *Ap. J.* 791:7

Bonfils X, Delfosse X, Udry S, et al. 2013. *Astron. Astrophys.* 549:A109

Borucki WJ, Agol E, Fressin F, et al. 2013. *Science* 340:587–90

Boyajian TS, von Braun K, van Belle G, et al. 2012. *Ap. J.* 757:112

Brack A, Horneck G, Cockell C, et al. 2010. *Astrobiology* 10(1):69–76

Broeg C, Fortier A, Ehrenreich D, et al. 2013. In *EPJ Web Conf.* 47:03005

Buccino AP, Lemarchand GA, Mauas PJ. 2006. *Icarus* 183:491

Burrows AS. 2014. *Nature* 513(7518):345–52

Cabrera J, Schneider J. 2007. *Astron. Astrophys.* 464:1133

Campante TL, Barclay T, Swift JJ. 2015. *Ap. J.* 799:2

Catanzarite JH, Shao M. 2011. *Ap. J.* 738:151

Charbonneau D, Brown TM, Latham DW, Mayor M. 2000. *Ap. J. Lett.* 529(1):L45–48

Chen EMA, Nimmo F, Glatzmaier GA. 2014. *Icarus* 229:11–30

Christensen PR, Pearl JC. 1997. *J. Geophys. Res.* 102:10875–80

Chyba CF, Hand KP. 2005. *Annu. Rev. Astron. Astrophys.* 43:31–74

Cincunegui C, D´ıaz RF, Mauas PJ. 2007. *Astron. Astrophys.* 461:1107–13

Clampin M, JWST Sci. Work. Group, JWST TransitWork. Group, et al. 2009. AST2010 Sci. White Pap., JamesWebb Space Telesc. ( JWST) Transit Sci.

Cnossen I, Sanz-Forcada J, Favata F, et al. 2007. *J. Geophys. Res.: Planets* 112:2008

Cockell CS. 1998. *J. Theor. Biol.* 193:717

Cockell CS. 2016. *Mol. Biol. Cell* 27(10):1553–55

Cockell CS, Kaltenegger L, Raven J. 2009. *Astrobiology* 9(7):623–36

Cohen O, Drake JJ, Glocer A, et al. 2014. *Ap. J.* 790:57

Cowan NB, Abbot DS, Voigt A. 2012. *Ap. J. Lett.* 752(1):L3

Cowan NB, Agol E, Meadows VS, et al. 2009. *Ap. J.* 700(2):915

Cowan NB, Strait TE. 2013. *Ap. J. Lett.* 765(1):L17

Crossfield IJM. 2015. *Publ. Astron. Soc. Pac.* 127(956):941–60

Cullum J, Stevens DP, Joshi MM. 2014. *Astrobiology* 14(8):645–50

Cullum J, Stevens DP, Joshi MM. 2016. *PNAS* 113(16):4278–83

Cuntz M. 2014. *Ap. J.* 780(1):14

Cuntz M, Guinan EF. 2016. *Ap. J.* 827:79

Danchi W, Lopez B. 2013. *Ap. J.* 769:27

Deming D, Seager S, Winn J, et al. 2009. *Publ. Astron. Soc. Pac.* 121:952–67

Demory B-O, Gillon M, Madhusudhan N, Queloz D. 2016. *Mon. Not. R. Astron. Soc.* 455(2):2018–27

Des Marais DJ, Harwit MO, Jucks KW, et al. 2002. *Astrobiology* 2(2):153–81

Dobos V, Turner EL. 2015. *Ap. J.* 804:41

Domagal-Goldman SD, Meadows VS, Claire MW, Kasting JF. 2011. *Astrobiology* 11:419–41

Domagal-Goldman SD, Segura A, Claire MW, et al. 2014. *Ap. J.* 792(2):90

Donahue TM, Hoffman JH, Hodges RR Jr., Waxson AJ. 1982. *Science* 216:630–33

Dressing CD, Charbonneau D. 2013. *Ap. J.* 767:95

Dressing CD, Charbonneau D. 2015. *Ap. J.* 807(1):45

Dumusque X, Pepe F, Lovis C, et al. 2012. *Nature* 1(5):4–8

Dvorak R, Pilat-Lohinger E, Bois E, et al. 2010. *Astrobiology* 10(1):33–43

Edson AR, Kasting JF, Pollard D, et al. 2012. *Astrobiology* 12:562–71

Edson AR, Lee S, Bannon P, et al. 2011. *Icarus* 212:1–13

Ehrenreich D, Tinetti G, Lecavelier des Etangs A, et al. 2006. *Astron. Astrophys.* 448:379

Elkins-Tanton LT, Seager S. 2008. *Ap. J.* 685:1237–46

Ferreira D, Marshall J, O'Gorman PA, Seager S. 2014. *Icarus* 243:236–48

Foley BJ. 2015. *Ap. J.* 812:36

Ford E, Seager S, Turner EL. 2001. *Nature* 412:885–87

Forgan D. 2014. *MNRAS* 437:1352

Forgan D, Kipping D. 2013. *MNRAS* 432:2994

Forget P, Pierrehumbert H. 1997. *Science* 278:1273–74

Fortier A, Beck T, Benz W, et al. 2014. In *Space Telescopes and Instrumentation 2014: Optical, Infrared, and Millimeter Wave*, ed. JM Oschmann Jr., M Clampin, GG Fazio, HA MacEwen. *Proc. SPIE* 9143:91432J.
Bellingham,WA, SPIE

France K, Froning CS, Linsky JL, et al. 2013. *Ap. J.* 763:149

Fu R, O'Connell RJ, Sasselov DD. 2010. *Ap. J.* 708:1326–34

Fujii Y, Kawahara H, Suto Y, et al. 2010. *Ap. J.* 715(2):866–80

Fujii Y, Kawahara H, Suto Y, et al. 2011. *Ap. J.* 738(2):184

Gaidos E. 2013. *Ap. J.* 770(2):90

Gaidos E, Williams DM. 2004. *New Astron.* 10(1):67–77
Garcıa Munoz A, Zapatero Osorio MR, Barrena R, et al. 2012. *Ap. J.* 755(2):103
Gardner JP, Mather JC, Clampin M, et al. 2006. *Space Sci. Rev.* 123:485
Gillon M, Jehin E, Lederer SM, et al. 2017. *Nature* 542:456–60
Goldblatt C, Robinson TD, Zahnle KJ, Crisp D. 2013. *Nat. Geosc.* 6:8
Goldblatt C,Watson AJ. 2012. *Philos. Trans. A Math. Phys. Eng. Sci.* 370(1974):4197–216
Goldblatt C, Zahnle K. 2011. *Clim. Past* 7:203–20
Gomez-Leal I, Pall´e E, Selsis F. 2012. *Ap. J.* 752(1):28
Grasset O, Schneider J, Sotin C. 2009. *Ap. J.* 693:722
Grenfell JL, Gebauer S, von Paris P, et al. 2011. *Icarus* 211(1):81–88
Grenfell JL, Rauer H, Selsis F, et al. 2010. *Astrobiology* 10(1):77–88
Grenfell JL, Stracke B, von Paris P, et al. 2007. *Planet Space Sci.* 55:661–71
Grießmeier J-M, Stadelmann A, Grenfell JL, et al. 2009. *Icarus* 199:526–35
Haghighipour N, Kaltenegger L. 2013. *Ap. J.* 777(2):172
Han C. 2008. *Ap. J.* 684:684
Hansen BMS. 2014. *Int. J. Astrobiol.* 14:267–78
Haqq-Misra JD, Saxena P, Wolf ET, Kopparapu RK. 2016. *Ap. J.* 827:2
Harman CE, Schwieterman EW, Schottelkotte JC, Kasting JF. 2015. *Ap. J.* 812(2):137
Hart MH. 1978. *Icarus* 33:23–39
Heath MJ, Doyle LR, Joshi MM, Haberle RM. 1999. 29:405
Hedelt P, von Paris P, Godolt M, et al. 2013. *Astron. Astrophys.* 553:A9
Hegde S, Kaltenegger L. 2013. *Astrobiology* 13(1):47–56
Hegde S, Paulino-Lima IG, Kent R, et al. 2015. *PNAS* 112(13):3886–91
Heller R, Pudritz R. 2015. *Ap. J.* 806:181
Heller R,Williams D, Kipping D, et al. 2014. *Astrobiology* 14:798
Henning WG, Hurford T. 2014. *Ap. J.* 789(1):30
Henning WG, O'Connell, Sasselov DD. 2009. *Ap. J.* 707(Scharf 2006):1000
Hinkel NR, Kane SR. 2013. *Ap. J.* 774:27
Holland HD. 2006. *Philos. Trans. R. Soc. B.* 361:903–15
Holman MJ, Murray NW. 2005. *Science* 307:5713
Hoyle F. 1958. In *Stellar Populations: Proc. Conf. Spons. Pontif. Acad. Sci. Vatican Obs.*, ed. DJK O'Connell, pp. 223–30. Amsterdam: North Holland
Hu R, Ehlmann BL, Seager S. 2012. *Ap. J.* 752(1):7
Hu R, Seager S, BainsW. 2013. *Ap. J.* 769:6
Huang SS. 1959. *Am. Sci.* 47:397
Joshi MM. 2003. *Astrobiology* 3:415–27
Joshi MM, Haberle RM. 2012. *Astrobiology* 12:3
Kadoya S, Tajika E. 2014. *Ap. J.* 790:107
Kaltenegger L. 2000. *Explor. Util. Moon* 462:199
Kaltenegger L. 2010. *Ap. J. Lett.* 712(2):L125–30
Kaltenegger L, Haghighipour N. 2013. *Ap. J.* 777(2):165
Kaltenegger L, Henning WG, Sasselov DD. 2010. *Astron. J.* 140:1370–80
Kaltenegger L, Sasselov DD. 2010. *Ap. J.* 708:1162–67
Kaltenegger L, Sasselov DD. 2011. *Ap. J. Lett.* 736(2):L25
Kaltenegger L, Sasselov DD, Rugheimer S. 2013. *Ap. J. Lett.* 775(2):L47
Kaltenegger L, Selsis F, Fridlund M, et al. 2009. *Astrobiology* 10(1):66
Kaltenegger L, Traub WA. 2009. *Ap. J.* 698:519

Kaltenegger L, Traub WA, Jucks KW. 2007. *Astron. J.* 658(1):598–616
Kane SR, Hill ML, Kasting JF, et al. 2016. *Ap. J.* 830(1):1
Kane SR, Hinkel NR. 2013. *Ap. J.* 762:7
Kasper ME, Beuzit J-L, Verinaud, et al. 2008. *Proc. SPIE* 7015:70151S
Kasting JF. 1988. *Icarus* 74(3):472–94
Kasting JF. 1992. *Orig. Life Evol. Biosph.* 20:199–231
Kasting JF. 2010. *Nature* 464(7289):687–89
Kasting JF, Catling D. 2003. *Annu. Rev. Astron. Astrophys.* 41:429–63
Kasting JF, Kopparapu R, Ramirez RM, Harman CE. 2014. *PNAS* 111(35):12641–46
Kasting JF, Whitmire DP, Reynolds RT. 1993. *Icarus* 101:108–28
Kerwin BA, Remmele RL. 2007. *J. Pharm. Sci.* 96:1468
Kharecha P, Kasting J, Siefert J. 2005. *Geobiology* 3(2):53–76
Kiang NY, Segura A, Tinetti G, et al. 2007a. *Astrobiology* 7:252–74
Kiang NY, Siefert J, Govindjee, Blankenship RE. 2007b. *Astrobiology* 7:222–51
Kipping DM. 2009. *MNRAS* 396:1797
Kipping DM, Bakos GÁ, Buchhave LA, Nesvorný D, Schmitt A. 2012. *Ap. J.* 750:115
Kipping DM, Fossey SJ, Campanella G. 2009. *MNRAS* 400:398
Kite ES, Manga M, Gaidos E. 2009. *Ap. J.* 700(2):1732–49
Kitzmann D. 2017. *Astron. Astrophys.* 600:A111
Kitzmann D, Alibert Y, Godolt M, et al. 2015. *Mon. Not. R. Astron. Soc.* 452(4):3752–58
Knutson HA, Charbonneau D, Allen LE, et al. 2007. *Nature* 447:183–85
Kopparapu RK. 2013. *Ap. J. Lett.* 767(1):L8
Kopparapu RK, Ramirez R, Kasting JF, et al. 2013. *Ap. J.* 765(2):131
Kopparapu RK, Ramirez RM, Schottel Kotte J, et al. 2014. *Ap. J. Lett.* 787(2):L29
Kopparapu RK, Wolf ET, Haqq-Misra J, et al. 2016. *Ap. J.* 819:84
Korenaga J. 2013. *Annu. Rev. Earth Planet Sci.* 41:117–51
Krissansen-Totton J, Schwieterman EW, Charnay B, et al. 2016. *Ap. J.* 817:31
Kruger M, Frenzel P, Conrad R. 2001. *Global Change Biol.* 7:49–63
Kuchner M. 2003. *Ap. J. Lett.* 596:L105
Lammer H, Kislyakova KG, Holmström M, Khodachenko ML, Grießmeier J-M. 2011. *Ap. Space Sci.* 335:9
Lammer H, Lichtenegger HI, Kulikov YN, et al. 2007. *Astrobiology* 7:185
Latham DW, Mazeh T, Stefanik RP, Mayor M, Burki G. 1989. *Nature* 339:38
Leconte J, Forget F, Charnay B, et al. 2013a. *Nature* 504(7479):268–71
Leconte J, Forget F, Charnay B, et al. 2013b. *Astron. Astrophys.* 554:A69
Leconte J, Wu H, Menou K, Murray N. 2015. *Science* 347(6222):632–35
Lederberg J. 1965. *Nature* 207(4992):9–13
Leger A, Grasset O, Fegley B, et al. 2011. *Icarus* 213(1):1–11
Leger A, Pirre M, Marceau FJ. 1993. *Astron. Astrophys.* 277:309–13
Leger A, Selsis F, Sotin C, et al. 2004. *Icarus* 169:499–504
Levi A, Sasselov D, Podolak M. 2013. *Ap. J.* 769(1):29
Levi A, Sasselov D, Podolak M. 2017. *Ap. J.* 838(1):24
Lewis KM, Sackett PD, Mardling RA. 2008. *Ap. J. Lett.* 685:L153
Lin HW, Abad GG, Loeb A. 2014. *Ap. J. Lett.* 792(1):L7
Lineweaver CH, Chopra A. 2012. *Annu. Rev. Earth Planet Sci.* 40:597–623
Linsenmeier M, Pascale S, Lucarini V. 2015. *Planet. Space Sci.* 105:43–59
Linsky JL, France K, Ayres T. 2013. *Ap. J.* 766:69

Lippincott ER, Eck RV, Dayhoff MO, Sagan C. 1967. *Ap. J.* 147:753
Lissauer JJ. 2007. *Ap. J. Lett.* 660(2):L149
Lissauer JJ, Barnes JW, Chambers JE. 2012. *Icarus* 217:77
Lissauer JJ, Fabrycky DC, Ford EB, et al. 2011. *Nature* 470(7332):53–58
Lloyd-Hart M, Angel R, Milton NM, et al. 2006. *Proc. SPIE* 6272:62720E
Lopez B, Schneider J, Danchi WC. 2005. *Ap. J.* 627:974–85
Lorenz RD, Lunine JI, McKay CP. 1997. *Geophys. Res. Lett.* 24(22):2905–8
Lovelock JE. 1965. *Nature* 207(997):568–70
Luger R, Barnes R. 2015. *Astrobiology* 15:119–43
Lundock R, Ichikawa T, Okita H, et al. 2009. *Astron. Astrophys.* 507:1649
Madden J, Kaltenegger L. 2017. *MNRS*. In press
Matsunaga T, Hieda K, Nikaido O. 1991. *Photochem. Photobiol.* 54:403
Mayor M, Queloz D. 1995. *Nature* 378:355–59
Meadows VS, Arney GN, Schwieterman EW, et al. 2016. *Astrobiology.* In review. arXiv:1608.08620
Menou K. 2015. *Earth Planet. Sci. Lett.* 429:20–24
Miguel Y, Kaltenegger L, Fegley B, Schaefer L. 2011. *Ap. J. Lett.* 742:L19
Mischna MA, Kasting JF, Pavlov AA, Freedman R. 2000. *Icarus* 145(2):546–54
Misra A, Krissansen-Totton J, Koehler MC, Sholes S. 2015. *Astrobiology* 15(6):462–77
Misra A, Meadows V, Claire M, Crisp D. 2014. *Astrobiology* 14:67
Mojzsis SJ, Arrhenius G, McKeegan KD, et al. 1996. *Nature* 384:55–59
Montañés-Rodríguez P, Pallé E, Goode PR. 2006. *Ap. J.* 651(1):544–52
Morton TD, Swift J. 2014. *Ap. J.* 791:10
Moskovitz NA, Gaidos E, Williams DM. 2009. *Astrobiology* 9:269
Noack L, Breuer D. 2014. *Planet Space Sci*. 98:41–49
Noack L, Hoening D, Rivoldini A, et al. 2016. *Icarus* 277:215–36
O'Malley-James JT, Kaltenegger L. 2017a. MNRAS In press. arXiv:1608.06930
O'Malley-James JT, Kaltenegger L. 2017b. *MNRS Lett.* 469:L26–30
O'Malley-James JT, Raven JA, Cockell CS, Greaves JS. 2012. 12(2):115–24
O'Neill C, Jellinek AM, Lenardic A. 2007. *Earth Planet Sci. Lett.* 261:20–32
Owen T. 1980. In *Strategies for the Search of Life in the Universe*, ed. MD Papagiannis, pp. 177–85. Dordrecht, Neth.: Reidel
Palle E, Ford EB, Seager S, et al. 2008. *Ap. J.* 676:1319–29
Palle E, Goode PR, Montañés-Rodríguez P, et al. 2016. *Geophys. Res. Lett.* 43(9):4531–38
Palle E, Osorio MRZ, Barrena R, et al. 2009. *Nature* 459:814–16
Pavlov AA, Kasting JF, Brown LL, et al. 2000. *J. Geophys. Res.* 105:11981–90
Petigura EA, Howard AW, Marcy GW, et al. 2013. *PNAS* 110:19273–78
Pierrehumbert RT. 2010. *Principles of Planetary Climate*. Cambridge, UK: Cambridge Univ. Press, 674
Pierrehumbert RT, Gaidos E. 2011. *Ap. J. Lett.* 734(1):L13
Pilcher CB. 2003. *Astrobiology* 3(3):471–86
Planavsky NJ, Reinhard CT, Wang X, et al. 2014. *Science* 346:635–38
Ramirez RM, Kaltenegger L. 2014. *Ap. J. Lett.* 797(2):L25
Ramirez RM, Kaltenegger L. 2016. *Ap. J.* 823(1):6
Ramirez RM, Kaltenegger L. 2017. *Ap. J. Lett.* 837:L4
Ramirez RM, Kopparapu R, Zugger ME, et al. 2014b. *Nat. Geosci.* 7(1):59–63
Ramirez RM, Kopparapu RK, Lindner V, Kasting JF. 2014a. *Astrobiology* 14:8
Ranjan S, Sasselov DD. 2016. *Astrobiology* 16:68
Rauer H, Catala C, Aerts C, et al. 2014. *Exp. Astron.* 38:249

Rauer H, Gebauer S, Paris PV, et al. 2011. *Astron. Astrophys.* 529(5):A8
Raymond SN, Scalo J, Meadows VS. 2007. *Ap. J.* 669:1
Read PL. 2014. *Astrobiology* 14(8):627–28
Reynolds RT, McKay CP, Kasting JF. 1987. *Adv. Space Res.* 7:125
Ribas I, Bolmont E, Selsis F, et al. 2016. *Astron. Astrophys.* 596:A111
Ribas I, Guinan EF, Gudel M, Audard M. 2005. *Ap. J.* 622:1
Ricker GR, Winn JN, Vanderspek R, et al. 2014. *Proc. SPIE* 9143:914320
Riedel AR, Murphy SJ, Henry TJ, et al. 2011. *Astronomical J.* 142(4):104
Ritson D, Sutherland JD. 2012. *Nat. Chem.* 4:895–99
Robinson TD, Ennico K, Meadows VS, et al. 2014. *Ap. J.* 787:171
Robinson TD, Meadows VS, Crisp D, et al. 2011. *Astrobiology* 11:393–408
Rodler F, Lopez-Morales M. 2014. *Ap. J.* 781:54
Rogers L. 2015. *Ap. J.* 801:41
Rogers L, Seager S. 2010. *Ap. J.* 716:1208
Rothschild LJ. 2008. *Philos. Trans. R. Soc. B* 363(1504):2787–801
Rothschild LJ, Mancinelli RL. 2001. *Nature* 409:1092–101
Rowe JF, Bryson ST, Marcy GW, et al. 2014. *Ap. J.* 784:45
Rugheimer S, Kaltenegger L, Segura A, et al. 2015a. *Ap. J.* 809:57
Rugheimer S, Kaltenegger L, Zsom A, et al. 2013. *Astrobiology* 13:251–69
Rugheimer S, Segura A, Kaltenegger L, et al. 2015b. *Ap. J.* 806(1):137
Rushby AJ, Claire MW, Osborn H, Watson AJ. 2013. *Astrobiology* 13(9):833–49
Sagan C, Mullen G. 1972. *Science* 177:52–56
Sagan C, Thompson WR, Carlson R, et al. 1993. *Nature* 365(6448):715–21
Sanromá E, Pallé E, Parenteau MN, et al. 2013. *Ap. J.* 780(1):52
Sartoretti P, Schneider J. 1999. *Astron. Astrophys. Suppl. Ser.* 134:553
Scalo J, Kaltenegger L, Segura A, et al. 2007. *Astrobiology* 7(1):85–166
Schaefer L, Fegley B Jr. 2009. *Ap. J.* 703:113
Schaefer L, Sasselov D. 2015. *Ap. J.* 801(1):40
Scharf CA. 2006. *Ap. J.* 648:1196
Schindler TL, Kasting JF. 2000. *Icarus* 145:262–71
Schwartz JC, Sekowski C, Haggard HM, et al. 2016. *Mon. Not. R. Astron. Soc.* 457(1):926–38
Schwieterman EW, Meadows VS, Domagal-Goldman SD, et al. 2016. *Ap. J. Lett.* 819(1):L13
Seager S, Bains W, Hu R. 2013. *Ap. J.* 777(2):95
Seager S, Bains W, Petkowski JJ. 2016. *Astrobiology* 16(6):465–85
Seager S, Deming D. 2010. *Annu. Rev. Astron. Astrophys.* 48:631–72
Seager S, Turner EL, Schafer J, Ford EB. 2005. *Astrobiology* 5:372–90
Segura A, Kasting JF, Meadows V, et al. 2005. *Astrobiology* 5:706–25
Segura A, Krelove K, Kasting JF, et al. 2003. *Astrobiology* 3(4):689–708
Segura A, Meadows VS, Kasting JF, et al. 2007. *Astron. Astrophys.* 472(2):665–79
Segura A, Walkowicz LM, Meadows V, et al. 2010. *Astrobiology* 10(7):751–71
Selsis F. 2000. In *Darwin and Astronomy—The Infrared Space Interferometer. Proc. Int. Symp., Stockholm, Sweden, Nov. 17–19, 1999*, ESA SP-451, pp. 133–42. Noordwijk, Neth.: ESA
Selsis F. 2004. In *Extrasolar Planets: Today and Tomorrow*, ed. J-P Beaulier, A Lecavelier des Etangs, C Terquen. *ASP Conf. Ser.* 321:170–82. San Francisco: ASP
Selsis F, Despois D, Parisot J. 2002. *Astron. Astrophys.* 388(3):985–1003
Selsis F, Kaltenegger L, Paillet J. 2008. *Phys. Scr.* T130:014032
Selsis F, Kasting JF, Levrard B, et al. 2007. *Astron. Astrophys.* 476:1373–87

Selsis F, Wordsworth R, Forget F. 2011. *Astron. Astrophys.* 532:14
Senanayake SD, Idriss H. 2006. *PNAS* 103(5):1194–98
Shields AL, Ballard S, Johnson JA. 2016. *Phys. Rep.* 663:1–38
Shields AL, Meadows VS, Bitz CM, et al. 2013. *Astrobiology* 13:715
Silburt A, Gaidos E, Wu Y. 2015. *Ap. J.* 799:180
Simon A, GyM Szabo, Szatmary K. 2009. *Earth Moon Planets* 105:385
Snellen I, de Kok R, Le Poole R, et al. 2013. *Ap. J.* 764:182
Southam G, Westall F. 2007. In *Treatise on Geophysics* 10:421–37
Spiegel DS, Menou K, Scharf CA. 2009. *Ap. J.* 691(1):596–610
Stamenkovic V, Noack L, Breuer D, Spohn T. 2012. *Ap. J.* 748:41
Stern SA. 2003. *Astrobiology* 3(2):317–21
Sullivan PW, Winn JN, Berta-Thompson ZK, et al. 2015. *Ap. J.* 809(1):77
Szabo GyM, Szatmary K, Diveki Z, Simon A. 2006. *Astron. Astrophys.* 450:395
Tackley PJ, Ammann M, Brodholt JP, et al. 2013. *Icarus* 225(1):50–61
Tevini M, ed. 1993. *UV-B Radiation and Ozone Depletion: Effects on Humans, Animals, Plants, Microorganisms, and Materials*. Boca Raton, FL: Lewis
Tian F. 2015. *Annu. Rev. Earth Planet. Sci.* 43:459–76
Tian F, France K, Linsky JL, et al. 2014. *Earth Planet Sci. Lett.* 385:22–27
Tian F, Ida S. 2015. *Nat. Geosci.* 8(3):177–80
Tinetti G, Rashby N, Yung Y. 2006. *Ap. J. Lett.* 644:L129–32
Traub WA. 2003. In *Scientific Frontiers in Research on Extrasolar Planets*, ed. D Deming, S Seager. *ASP Conf. Ser.* 294:595–602. San Francisco: ASP
Traub WA. 2011. *Ap. J.* 745:20
Traub WA. 2016. *Ap. J.* Submitted. arXiv:1605.02255
Traub WA, Jucks K. 2002. In *Atmospheres in the Solar System: Comparative Aeronomy*, Vol. 130, *Geophys. Monogr. Ser.*, ed. M Mendillo, A Nagy, JH Waite, pp. 369–80. Washington, DC: Am. Geophys. Union
Trenberth KE, Fasullo JT. 2009. *Meteorol. Z.* 18(4):369–77
Turbet M, Leconte J, Selsis F, et al. 2016. *Astron. Astrophys.* 596:A112
Turnbull MC, Traub WA, Jucks KW, et al. 2006. *Ap. J.* 644:551–59
Udry S, Santos NC. 2007. *Annu. Rev. Astron. Astrophys.* 45:397–439
Valencia D, O'Connell RJ. 2009. *Earth Planet. Sci. Lett.* 286(3–4):492–502
Valencia D, O'Connell RJ, Sasselov DD. 2007. *Ap. J. Lett.* 670:L45–48
van Heck HJ, Tackley PJ. 2011. *Earth Planet Sci. Lett.* 310:252–61
Vidal-Madjar A, Arnold A, Ehrenreich D, et al. 2010. *Astron. Astrophys.* 523:A57
Villaver E, Livio M. 2007. *Ap. J.* 661:1192–201
Vladilo G, Murante G, Silva L, et al. 2013. *Ap. J.* 767:65
Voet D, Gratzer WB, Cox RA, Doty P. 1963. *Biopolymers* 1:193
von Braun K, Boyajian TS, Kane SR, et al. 2011. *Ap. J. Lett.* 729(2):L26
von Paris P, Selsis F, Kitzmann D, Rauer H. 2013. *Astrobiology* 13:899
Walker JCG. 1977. *Evolution of the Atmosphere*. New York: Macmillan
Walker JCG, Hays PB, Kasting JF. 1981. *J. Geophys. Res.* 86:9776–82
Ward WR. 1982. *Icarus* 50:444–48
Weiss LM, Marcy GW. 2014. *Ap. J. Lett.* 783(1):L6
West AA, Morgan DP, Bochanski JJ, et al. 2011. *Astronom. J.* 141:97
Williams DM, Gaidos E. 2008. *Icarus* 195(2):927–37
Williams DM, Kasting J. 1997. *Icarus* 129:254–67

Williams DM, Pollard D. 2002. In *The Evolving Sun and Its Influence on Planetary Environments*, ed. B Montesinos, A Gimenez, EF Guinan. *ASP Conf. Ser.* 269:201–13. San Francisco: ASP

Williams JP, Cieza LA. 2011. *Annu. Rev. Astron. Astrophys.* 49:67–117

Winn JN, Fabrycky DC. 2015. *Annu. Rev. Astron. Astrophys.* 53:409–77

Wolf ET, Toon OB. 2014. *Geophys. Res. Lett.* 41:167–72

Wolf ET, Toon OB. 2015. *J. Geophys. Res.: Atmospheres* 120:5775

Wolfgang A, Rogers LA, Ford EB. 2016. *Ap. J.* 825(1):19

Woolf NJ, Smith PS, TraubWA, Jucks KW. 2002. *Ap. J.* 574:430–42

Wordsworth R. 2012. *Icarus* 219(1):267–73

Wordsworth R, Kalugina Y, Lokshtanov S, et al. 2017. *Geophys. Res. Lett.* 44:665–71

Wordsworth R, Pierrehumbert R. 2014. *Ap. J. Lett.* 785:L20

Wordsworth RD, Pierrehumbert RT. 2013. *Ap. J.* 778:154

Wright SA, Barton EJ, Larkin JE, et al. 2010. In *Groundbased and Airborne Instrumentation for Astronomy III*, ed. IS McLean, SK Ramsay, H Takami. *Proc. SPIE Conf. Ser.* 7735:77357P. Bellingham,WA: SPIE

Yang J, Boue G, Fabrycky DC, Abbot DS. 2014. *Ap. J. Lett.* 787(1):L2

Yang J, Cowan NB, Abbot DS. 2013. *Ap. J. Lett.* 771:L45

Youngblood A, France K, Parke Loyd RO, et al. 2016. *Ap. J.* 824:101

Zahnle K, Arndt N, Cockell C, et al. 2007. *Space Sci. Rev.* 129(1–3):35–78

Zahnle K, Haberle RM, Catling DC, Kasting JF. 2008. *J. Geophys. Res. Planets* 113:E11004

Zendejas J, Segura A, Raga AC. 2010. *Icarus* 210:539–44

Zeng L, Sasselov DD. 2013. *Publ. Astron. Soc. Pac.* 125(925):227–39

Zeng L, Sasselov DD, Jacobsen SB. 2016. *Ap. J.* 819(2):127

Zsom A. 2015. *Ap. J.* 813:9

Zsom A, Kaltenegger L, Goldblatt C. 2012. *Icarus* 221(2):603–16

Zsom A, Seager S, deWit J, Stamenkovic V. 2014. *Ap. J.* 778:109